\documentclass[pdflatex,sn-mathphys-num]{sn-jnl}

\usepackage{graphicx}%
\usepackage{multirow}%
\usepackage{amsmath,amssymb,amsfonts}%
\usepackage{amsthm}%
\usepackage{mathrsfs}%
\usepackage[title]{appendix}%
\usepackage{xcolor}%
\usepackage{textcomp}%
\usepackage{manyfoot}%
\usepackage{booktabs}%
\usepackage{algorithm}%
\usepackage{algorithmicx}%
\usepackage{algpseudocode}%
\usepackage{listings}%
\usepackage{caption}
\usepackage{subcaption}
\usepackage{hyperref}

\begin{document}

\title[B C Roy, A Beesham and B C Paul]{Dynamics  of late time universe in $f(Q)$ gravity}


\author[1]{\fnm{Bikash Chandra} \sur{Roy}}\email{bcroy.bcr25@gmail.com}

\author[2]{\fnm{Aroonkumar} \sur{Beesham}}\email{abeesham@yahoo.com}

\author*[1]{\fnm{Bikash Chandra} \sur{Paul}}\email{bcpaul@nbu.ac.in}

\affil[1]{\orgdiv{Department of Physics}, \orgname{University of North Bengal}, \orgaddress{\street{Raja Rammohunpur, Darjeeling}, \city{Siliguri}, \postcode{734013}, \state{West Bengal}, \country{India}}}

\affil[2]{\orgdiv{Faculty of Natural Sciences}, \orgname{Mangosuthu University of Technology}, 
\city{Umlazi 4066}, 
\country{South Africa}}



\abstract{We construct cosmological model  in  nonmetricity scalar functional gravitational Lagrangian $f(Q)$ which describes the dynamical evolution of the late accelerating universe. Cosmological models are constructed considering different functional of $f(Q)$ gravity where $Q$ in the gravitational action. We obtain cosmological model probing late universe with a constant jerk parameter. The observational constraints that are imposed on the model parameters for a realistic scenario estimated using the observational Hubble data and the Pantheon dataset. The evolution of the deceleration parameter, energy density and the equation of state (EoS) parameter are also explored. The transition of the universe from a deceleration to an accelerating phase is investigated in different framework of $f(Q)$ theories. We also analyzed the variation of the effective EoS parameter and found that  the matter content in the universe  favours quintessence type fluid in all the $f(Q)$-gravity. The energy conditions for a realistic scenario are examined and noted that the effective fluid violates the strong energy condition.}




\maketitle

\section{Introduction}
\label{sec:intro}

In modern cosmology it is accepted that the universe emerged out from an inflationary phase in the past entered into matter dominated phase and thereafter at a later phase entered into accelerating universe. The prediction of late acceleration originated analyzing the recent advances in observational cosmology namely the type Ia supernovae \cite{1999ApJ...517..565P, riess_observational_1998, riess2004type}, Cosmic Microwave Background Radiation \cite{Spergel_2003} and large-scale structure observations \cite{Koivisto_2006,Daniel_2008,Minami_2020}. Although the early universe is understood with homogeneous scalar field in particle physics,  the late accelerating phase cannot be understood with fields in the standard model of particle physics. Inspite of the fact that the General relativity (GR) is the most successful fundamental gravitational theory which describes the large-scale structure of the universe,  it fails to accommodate the late universe behaviour unless one considers a theory beyond GR or exotic matter in the matter sector. Initially the late acceleration is examined introducing a cosmological constant term in GR. Such cosmological model is known as the Lambda cold dark matter or $\Lambda$CDM model, which yields the late-time cosmic acceleration of the universe that was found to fit well with the observations. But in the $\Lambda$CDM model conceptual issues namely, fine-tuning and cosmic coincidence problems \cite{SAHNI_2000, Carroll_2001, Padmanabhan_2003, Peebles_2003} emerged. Consequently, in the literature alternative cosmological model is explored without $\Lambda$. A spurt in research activities is found with a modified theory of gravity beyond GR that accommodate late acceleration.

Since normal matter in GR fails to accommodate the late acceleration of the universe,  the matter sector of the EFE with different dynamical Dark Energy (DE) models , namely, models consisting of one or more scalar fields namely, quintessence \cite{chiba_quintessence_1999, amendola_coupled_2000, martin_quintessence_2008}, Chaplygin gas \cite{Kamenshchik_2001} and its variations \cite{Bento_2002, Benaoum_2002} are considered. Alternatively, there are proposals to modify the gravitational sector with different gravitational theories, namely, $f(R)$ theories of gravity \cite{sotiriou_fr_2010,De_Felice_2010} where $R$ is the Ricci scalar, $f(R, \mathcal{T})$ gravity \cite{Harko_2011} with $\mathcal{T}$ being the trace of the energy-momentum tensor, modified Gauss-Bonnet gravity \cite{NOJIRI20051,PhysRevD.73.084007, PhysRevD.76.044027}, Brane world gravity \cite{Maartens_2010,Brax_2004}, Horava-Lifshitz theory of gravity \cite{wang_horava_2017}. In the literature \cite{starobinsky198030,PhysRevD.78.063504,Capozziello_2012_AdP,Myrzakulov_2012,RUDRA2021115428,Paul_2022,Nojiri_2006,Di_Valentino_2023,Zhang_2020} different modified gravitational theories are considered to describe several issues in astrophysical and cosmological observations and viability of the models are also examined using astronomical observations.

Gravity based on torsion is considered to obtain cosmological model in recent times. The torsion is the field describing gravity, called the teleparallel equivalent of general relativity (TEGR) \cite{pellegrini1963tetrad,PhysRevD.19.3524}. In TEGR, the curvature and non-metricity are zero and the Weitzenböck connection is the affine connection \cite{Linder_2010,Maluf_2013,Aldrovandi:2013wha}. The fundamental objects are tetrads by which one can derive the affine connection, the torsion invariants, and finally the field equations. TEGR is also modified further similar to the modification of Einstein Hilbert (EH) action  based on curvature formulation (the action that yields the Einstein field equations). The simplest modification is the $f(T)$ gravity \cite{Capozziello_2011}, where $T$ is the torsion scalar. An important advantage of the $f(T)$ gravity theory is that the field equations are of second order, unlike in $f(R)$ gravity, which in the metric approach is a fourth order theory. $f(T)$ gravity theories have been widely applied to the study of astrophysical objects and cosmic observations, and in particular, they are extensively used to explain the late-time accelerating expansion of the Universe without the need to introduce DE \cite{PhysRevD.75.084031,ferraro_born-infeld_2008,bengochea_dark_2009,capozziello2017constraining, Awad_2018,jimenez2018teleparallel}. Chaudhary {\it et al.} obtained cosmological models with two different forms of DE, namely modified Chaplygin-Jacobi (MCJ) and modified Chaplygin-Abel (MCA) gases, in the context of $f(T)$ gravity in a non-flat FLRW universe \cite{chaudhary2024_MCG}. The nature of the DE in the context of $f(T)$ gravity  (a detailed discussion of teleparallel theories see \cite{Cai_2016}) is explored for the evolution of the late  universe.

In recent times, another equivalent formulation of GR based on non-metricity in which a flat spacetime is considered in cosmology with total curvature and vanishing torsion. This formulation is called the symmetric teleparallel equivalent of GR (STEGR) \cite{nester1999symmetric, adak2004solution, ADAK_2006, ADAK_2013}. Similar to $f(R)$ theories, an extension of STEGR is $f(Q)$ gravity \cite{Jim_nez_2018}, where modifications based on a Lagrangian density which is a function of the non-metricity scalar $Q$. However when $f(Q) = Q$ one recovers STEGR. The class of theories obtained from the generalized Friedmann equation is found to accommodate the accelerated expansion of the universe. The modified $f(Q)$ gravity is studied phenomenologically in the literature \cite{Dialektopoulos_2019, PhysRevD.101.103507,bajardi2020bouncing,PhysRevD.102.124029,mandal2020energy,Arora_2022}. It is shown that the $f(Q)$ theory is one of the most promising alternative gravity formulations for interpreting cosmological observations\cite{lu2019_1906.08920, lazkoz2019observational, anagnostopoulos2021first, Narawade_2023, Atayde_2021}. The scalar, vector, and tensor modes of linear perturbations are studied in $f(Q)$ gravity \cite{PhysRevD.101.103507}, and the modified gravity is found to support the constraints imposed by Big Bang Nucleosynthesis (BBN) \cite{anagnostopoulos2023new}. Recently, $f(Q)$ gravity is taken up \cite{Dimakis_2021} to accommodate the late-time acceleration of the universe and DE (DE) \cite{lazkoz2019observational, anagnostopoulos2021first, koussour2022late, Solanki_2022}. In \cite{rana_2024a}, a bulk viscous cosmological model is studied in the framework of $f(Q)$ gravity, where it  is shown that the cosmological model favours quintessence-type behaviour which can describe the late-time scenario. The dynamical system analysis is used to study cosmological dynamics in $f(Q)$ theory at the background and perturbative levels \cite{khyllep2021cosmological, PhysRevD.107.044022, Narawade_2023_a3}. Recently, $f(Q)$ theories with $Q$ coupled non-minimally to the matter Lagrangian and the trace of the stress-energy tensor are developed to study the evolutionary dynamics \cite{Harko_2018, xu2019f}.

Motivated by the exciting features of the $f(Q)$ gravity, we investigate the late-time acceleration of a spatially flat universe with a constant cosmological perturbation namely, jerk parameter in the present paper and estimate the observational constraints of model parameters. We estimate the observational constraints on the the cosmological model parameters for a viable scenario making use of the observational data. We consider three different forms of $f(Q)$: (i) the power-law model, $f(Q) = Q + \alpha\left(\frac{Q}{Q_{0}}\right)^{n}$, where $\alpha$ and $n$ are the free parameters and $Q_{0}$ is the present value of non-metricity scalar $Q$ \cite{PhysRevD.101.103507}, (ii) $f(Q) = Q + nQ_{0}\sqrt{\frac{Q}{\lambda Q_{0}}}\ln\left(\frac{\lambda Q_{0}}{Q}\right)$, where $n$ and $\lambda$ are free parameters \cite{anagnostopoulos2023new} and (iii) $f(Q) = Qe^{\beta\frac{Q_{0}}{Q}}$, where $\beta$ is the model parameter \cite{anagnostopoulos2021first}. In the paper we consider a universe with a constant jerk parameter $j$ \cite{Pradhan_2023},  the concept of jerk parameters in cosmology provides valuable insights into the dynamics of the universe and the nature of DE. The fourth term of the Taylor series expansion of the scale factor $a(t)$ around the present time $t_{0}$ represents the jerk parameter. Mathematically, we obtain for the scale factor of the universe at $t$, which is Taylor expanded about $t_{0}$ is
\begin{equation}
    \frac{a(t)}{a(t_{0})} = 1 + H_{0}(t - t_{0}) - \frac{1}{2!}q_{0}H_{0}^{2}(t-t_{0})^{2} + \frac{1}{3!}j_{0}H_{0}^{3}(t-t_{0})^{3} + O[(t -t_{0})^{4}],
\end{equation}
where 
\begin{equation}
    H_{0} \equiv \left(\frac{1}{a}\frac{da}{dt}\right)_{t = t_{0}},\;\;\; q_{0} \equiv - \left(\frac{1}{aH^{2}}\frac{d^2a}{dt^2}\right)_{t = t_{0}},\;\;\; j_{0} \equiv \left(\frac{1}{aH^{3}}\frac{d^3a}{dt^3}\right)_{t = t_{0}}
\end{equation}
represent the present-day values of the Hubble parameter, deceleration parameter, and jerk parameter, respectively. It is known that the jerk parameter ($j$) is useful for studying the instability of a cosmological model along with the state finder diagnostic \cite{Sahni_2003}. The deviations from the standard cosmological model can be tested from the evolutionary feature of $j$. Alternatively, if one assumes  a constant $j$, it yields the Hubble parameter that  can be used for a good fit with the observed data. In the paper, the latest compilation of 31 Hubble data (HD) measured by the differential age (DA) method ( also known as cosmic chronometer (CC)) in the redshift range $0.07 \leq z \leq 1.965$  \cite{Sharov_2018} and 1048 Pantheon data set of apparent magnitude in the redshift range $0.01 \leq z \leq 2.26$ \cite{Scolnic_2018} are used to study the confrontation of the cosmological model with the observation, which however, determines the constraints on the model parameters. The best-fit present values of the jerk parameter ($j_{0}$), the deceleration parameter ($q_{0}$), and the Hubble parameter ($H_{0}$) are explored using the CC and a joint analysis of CC and the Pantheon data set. The evolution of the cosmological parameters, specifically the energy density and the equation of state (EoS) parameters for the cosmological models are also tested. The viability of these models is probed using energy conditions to ensure consistency with observation for a modified gravity $f(Q)$.

The paper is organized as follows: In section \ref{sec:FE} the basic field equations are given. As we do not know suitable $f(Q)$ we consider three different models to study the late universe in section \ref{cosmo:models}. In section \ref{sec:HP}, the Hubble parameter is estimated using the constant value of the jerk parameter. Using cosmic chronometers (CC), combined CC + Pantheon datasets the present value of Hubble parameter $H_{0}$, the jerk parameter $j_{0}$ and the deceleration parameter $q_{0}$ are constraints in the section \ref{sec: obs}. In section \ref{sec:evo}, the evolution of cosmological parameters are studied. The energy densities and EoS parameters of DE and the effective fluids for different $f(Q)$ models are described in subsection \ref{sec:edseos} and all the energy conditions for the validation of the models are studied in section \ref{sec:ECs} . Finally, in section \ref{sec:cls}, we give a brief discussion.

\section{Field Equations}
\label{sec:FE}
The general action in $f(Q)$ gravity is \cite{Jim_nez_2018}
\begin{equation}\label{FE:1}
S = - \frac{1}{2}\int f(Q)\sqrt{-g}\; d^{4}x + \int\mathcal{L}_{m}\sqrt{-g}d^{4}x,
\end{equation}
where $f(Q)$ is an arbitrary function of the non-metricity scalar $Q$, $g$ is the determinant of the matric $g_{\mu\nu}$ and $\mathcal{L}_{m}$ is the matter Lagrangian density. In this case, the fundamental object is the non-metricity tensor which defined by
\begin{equation}\label{FE:2}
    Q_{\alpha\mu\nu} = \nabla_{\alpha}g_{\mu\nu}.
\end{equation}
From the non-metricity tensor $Q_{\alpha\mu\nu}$, the disformation can be defined as
\begin{equation}\label{FE:3}
    L^{\alpha}_{\mu\nu} = \frac{1}{2}Q^{\alpha}_{\mu\nu} - Q_{(\mu\nu)}^{\alpha},
\end{equation}
which measures the separation of the Levi-Civita connection the symmetric part of the full connection with it. The non-metricity conjugate is defined as
\begin{equation}\label{FE:4}
P^{\alpha}{}_{\mu\nu} = -\frac{1}{2}L^{\alpha}{}_{\mu\nu} + \frac{1}{4}(Q^{\alpha} - \tilde{Q}^{\alpha})g_{\mu\nu} - \frac{1}{4}\delta^{\alpha}_{(\mu}Q_{\nu)},
\end{equation}
where $Q_{\alpha} \equiv Q_{\alpha}^{\;\;\mu}{}_{\mu}$ and $\tilde{Q}_{\alpha} \equiv Q^{\mu}_{\;\;\alpha\mu}$ are the trace of the non-metricity tensor. Therefore, the non-metricity scalar that will play a central role in $f(Q)$ gravity is defined as
\begin{equation}\label{FE:5}
    Q = - Q_{\alpha\mu\nu}P^{\alpha\mu\nu}.
\end{equation}

Variation of the action (\ref{FE:1}) with respect to components of the metric tensor leads to the modified field equations for $f(Q)$ gravity which is given by
\begin{equation}\label{FE:6}
\frac{2}{\sqrt{-g}}\nabla_{\alpha}\left(\sqrt{-g}f_{Q}P^{\alpha}{} _{\mu\nu}\right) + \frac{1}{2}g_{\mu\nu}f + f_{Q}\left(P_{\mu\alpha\beta}Q_{\nu}{}^{\alpha\beta} - 2Q_{\alpha\beta\mu}P^{\alpha\beta}_{\nu}\right) = T_{\mu\nu},
\end{equation}
where $f_{Q} \equiv \frac{\partial f(Q)}{\partial Q}$ and $T_{\mu\nu} = - \frac{2}{\sqrt{- g}}\frac{\delta(\sqrt{-g}\mathcal{L}_{m})}{\delta g^{\mu\nu}}$ is the energy-momentum tensor. We consider matter described by a perfect fluid whose energy-momentum tensor is,
\begin{equation}\label{FE:7}
    T_{\mu\nu} = (\rho + p)u_{\mu}u_{\nu} + pg_{\mu\nu},
\end{equation}
where $u_{\mu}$ is the four-velocity satisfying $u_{\mu}u^{\mu} = -1$, $\rho$ and $p$ are the energy density and pressure of the perfect fluid, respectively. 

We consider a homogeneous, isotropic and spatially flat Friedmann-Lema?tre-Robertson-Walker (FLRW) universe, which is given by
\begin{equation}\label{FE:8}
ds^{2} = - dt^{2} + a^{2}(t)\delta_{\mu\nu}dx^{\mu}dx^{\nu},    
\end{equation}
where $a(t)$ denotes the scale factor of the universe. The non-metricity scalar is given by $Q = 6H^{2}$, where $H = \frac{\dot{a}}{a}$ is the Hubble parameter and the upper dot denotes derivative with respect to cosmic time $t$. The modified gravity is described by $f(Q) = Q + F(Q)$, where $F(Q)$ is a non-linear function of $Q$. Using the FLRW metric, the corresponding Friedman equations become
\begin{equation}\label{FE:9}
3H^{2} = \rho + \frac{F}{2} - QF_{Q},
\end{equation} 
\begin{equation}\label{FE:10}
2\left(2QF_{QQ} + F_{Q} + 1\right)\dot{H} + \frac{1}{2}\left(Q + 2QF_{Q} - F\right) = - p,
\end{equation}
where $F_{Q} \equiv \frac{dF}{dQ}$ and $F_{QQ} \equiv \frac{d^{2}F}{dQ^{2}}$. Also, the matter fluid satisfies the conservation equation
\begin{equation}\label{FE:11}
    \dot{\rho} + 3H(1 + \omega)\rho = 0,
\end{equation}
with $\omega \equiv \frac{p}{\rho}$ the matter equation of state (EoS) parameter. First, we consider a universe filled with pressureless matter and radiation fluids, therefore we can write the total energy density and pressure as
\begin{equation}\label{FE:12}
    \rho = \rho_{m} + \rho_{r};\hspace{2cm} p = \frac{1}{3}p_{r},
\end{equation}
where $\rho_{m}$ and $\rho_{r}$ denote the energy density for pressureless matter and radiation, respectively. The above equations (\ref{FE:9}) and (\ref{FE:10}) can be expressed as
\begin{equation}\label{FE:13}
    3H^{2} = \rho_{m} + \rho_{r} +\rho_{DE},
\end{equation}
\begin{equation}\label{FE:14}
    2\dot{H} + 3H^{2} = - \frac{1}{3}p_{r} - p_{DE},
\end{equation}
where $\rho_{DE}$ and $p_{DE}$ are the DE density and pressure, which can be expressed as
\begin{equation}\label{FE:15}
    \rho_{DE} = \frac{F}{2} - QF_{Q},
\end{equation}
\begin{equation}\label{FE:16}
    p_{DE} = 2\left(2QF_{QQ} + F_{Q}\right)\dot{H} - \rho_{DE}.
\end{equation}
Therefore, the EoS parameter of DE is given by
\begin{equation}\label{FE:17}
    \omega_{DE} = \frac{p_{DE}}{\rho_{DE}} = - 1 + \frac{4(2QF_{QQ} + F_{Q})}{F - 2QF_{Q}}.
\end{equation}
The EoS parameter $\omega_{DE}$ for DE provided (i) the $\Lambda$CDM model for $\omega_{DE} = - 1$, (ii) a quintessence model for $ - 1 < \omega_{DE} < - \frac{1}{3}$, (iii) phantom model for $\omega_{DE} < - 1$.

Equations (\ref{FE:13}) and (\ref{FE:14}) yields the effective energy density $\rho_{eff}$ and effective pressure $p_{eff}$ of the total fluids which are
\begin{equation}\label{FE:18}
    \rho_{eff} = \rho_{m} + \rho_{r} + \frac{F}{2} - QF_{Q}
\end{equation}
and
\begin{equation}\label{FE:19}
    p_{eff} = \frac{1}{3}\rho_{r} + \left(QF_{Q} - \frac{F}{2}\right) - \left(\rho_{m} + \frac{4}{3}\rho_{r}\right)\left(\frac{2QF_{QQ} + F_{Q}}{2QF_{QQ} + F_{Q} + 1}\right).
\end{equation}
 Thus, the effective EoS parameter $\omega_{eff}$ becomes
\begin{equation}\label{FE:20}
    \omega_{eff} = \frac{p_{eff}}{\rho_{eff}} = - 1 + \frac{\Omega_{m} + \frac{4}{3}\Omega_{r}}{2QF_{QQ} + F_{Q} + 1},
\end{equation}
where $\Omega_{m} \equiv \frac{\rho_{m}}{3H^{2}}$ and $\Omega_{r} \equiv \frac{\rho_{r}}{3H^{2}}$ are the energy density parameters for pressureless matter and radiation, respectively. The EoS parameter $\omega_{eff}$ represents the matter-dominated if $\omega_{eff} = 0$, whereas radiation-dominated phase if $\omega_{eff} = \frac{1}{3}$. It is also helpful in distinguishing a decelerating universe from an accelerating one. For an accelerated universe one requires $\omega_{eff} < - \frac{1}{3}$. Thus the EoS parameter classifies three possible states for the accelerating universe which are (i) quintessence $(-1 < \omega_{eff} < - \frac{1}{3})$, (ii) phantom ($\omega_{eff} < -1$) and (iii) the cosmological constant ($\omega = -1$). The recent observational data employed to estimate constraint on the present value of the EoS parameter: $\omega_{eff} = - 1.29_{-0.12}^{+0.15}$ \cite{DIVALENTINO2016242}, Supernovae Cosmology Project, $\omega_{eff} = -1.035^{+0.055}_{-0.059}$ \cite{amanullah2010spectra}, Plank 2018, $\omega_{eff} = - 1.03 \pm 0.03$ \cite{planck_2018_results}. For a non-interacting  pressureless matter and radiation we obtain the following eqs. (\ref{FE:11}) : 
\begin{equation}\label{FE:21}
    \dot{\rho_{m}} + 3H\rho_{m} = 0, \;\;\;\;\dot{\rho_{r}} + 4H\rho_{r} = 0.
\end{equation}
On integrating the above equation, we get $\rho_{m} = \rho_{m0}a^{-3}$ and $\rho_{r} = \rho_{r0}a^{-4}$, where $\rho_{m0}$ and $\rho_{r0}$ denote the matter and radiation energy density at the present time.

The system of field equations (\ref{FE:15}) - (\ref{FE:17}) and (\ref{FE:18}) - (\ref{FE:20}) are non-linear which are taken up to investigate the dynamical evolution of DE and the effective fluids of the universe for a specific form of $F(Q)$. In the next section, we consider different $F(Q)$ forms and use them to explore the behaviour of the late Universe.

\section{Cosmological models in {\ensuremath{f(Q)}} gravity}
\label{cosmo:models}
In this section, we consider three different functions of $f(Q) = Q + F(Q)$ gravity to probe the late time evolution of the universe, which are as follows:
\subsection{Model-I}
In this case we  consider power-law model of $F(Q)$: \cite{Jim_nez_2018,PhysRevD.101.103507,PhysRevD.107.044022} as,
\begin{equation}\label{M1}
F(Q) = \alpha\left(\frac{Q}{Q_{0}}\right)^{n},
\end{equation} 
where $\alpha$ and $n$ are constant parameters and $Q_{0} = 6H_{0}^{2}$ is the present value of $Q$. For $n = 0$, the $F(Q)$ model is equivalent to the cosmological constant, while for $n = 1$ the model reduces to the symmetric teleparallel equivalent of GR (STEGR) \cite{Jim_nez_2018}. The power-law model is relevant for the early universe description when $n > 1$, however, the late accelerating scenario can be realized when $n < 1$. When $n = - 1$, the theory is confronted with late universe observations \cite{PhysRevD.103.063505}. For the constraints $0 < n < 1$, the evolutionary feature of the universe in $f(Q)$ gravity can be studied using autonomous differential equations obtained from the field equations, afterwards we investigate those solutions using perturbations outlined for its dynamical behaviour \cite{PhysRevD.107.044022}.

From equations (\ref{FE:15}) and (\ref{FE:16}), we determine the DE (DE) density and pressure as,
\begin{equation}\label{Eq:M1rde}
    \rho_{DE}(z) = \frac{\alpha(1 - 2n)}{2}\left(\frac{H^{2}(z)}{H_{0}^{2}}\right)^{n}, 
\end{equation}
\begin{equation}\label{Eq:M1pde}
    p_{DE}(z) = \frac{\alpha n (1 - 2n)}{3}\left(\frac{H^{2}(z)}{H_{0}^{2}}\right)^{n}\frac{(1 + z)H'(z)}{H(z)} - \rho_{DE}(z),
\end{equation}
where prime denotes differentiation with respect to $z$. Using the above expressions, the EoS parameter of DE of the form
\begin{equation}\label{Eq:M1wde}
    \omega_{DE}(z) = - 1 + \frac{2n(1 + z)}{3}\frac{H'(z)}{H(z)}.
\end{equation}
The effective energy density and pressure given by equations (\ref{FE:18}) and  (\ref{FE:19}) can be expressed in terms of the redshift parameter ($z$) which yield:
\begin{equation}\label{Eq:M1reff}
    \rho_{eff}(z) = 3H_{0}^{2}\left[\Omega_{m0}(1 + z)^{3} + \Omega_{r0}(1 + z)^{4}\right] + \frac{\alpha(1 - 2n)}{2}\left(\frac{H^{2}(z)}{H_{0}^{2}}\right)^{n},
\end{equation}
\[p_{eff}(z) = H_{0}^{2}\Omega_{r0}(1 + z)^{4} + \left[- 1 + \frac{2n(1 + z)}{3}\frac{H'(z)}{H(z)}\right]\]
\begin{equation}\label{Eq:M1peff}
    \hspace{3cm}\times \frac{\alpha (1 - 2n)}{2}\left(\frac{H^{2}(z)}{H_{0}^{2}}\right)^{n}.
\end{equation}
The effective EoS parameter $\omega_{eff}$ for the power-law model is obtained from equations (\ref{Eq:M1reff}) and (\ref{Eq:M1peff}). In the above, there are two parameters namely, $\alpha$ and $n$; one of them can be eliminated using the present value of the cosmological parameters. Using equations (\ref{FE:9}) and (\ref{M1}), the parameter $\alpha$ becomes
\begin{equation}\label{Eq:alpha}
    \alpha = \frac{6H_{0}^{2}(1 - \Omega_{m0} - \Omega_{r0})}{1 - 2n}.
\end{equation}
For $n = \frac{1}{2}$, $\alpha$ becomes infinity which is not acceptable.
\subsection{Model-II} 
In this case we consider a Log-square-root model of $F(Q)$ which is given by \cite{anagnostopoulos2023new},
\begin{equation}\label{M2}
   F(Q) = nQ_{0}\sqrt{\frac{Q}{\lambda Q_{0}}}\ln{\left(\frac{\lambda Q_{0}}{Q}\right)},
\end{equation}
where $n$ and $\lambda$ are free parameters, and $Q_{0} = 6H_{0}^{2}$ with $H_{0}$ the present value of the Hubble parameter and $\lambda > 0$. Analyzing the model with observations we found that the model passes through the BBN constraints \cite{anagnostopoulos2023new}. In this case, the DE density and pressure are given by
\begin{equation}\label{Eq:M2rde}
    \rho_{DE}(z) = \frac{6nH^{2}(z)}{\lambda\sqrt{\frac{H^{2}(z)}{\lambda H_{0}^{2}}}},
\end{equation}
\begin{equation}\label{Eq:M2pde}
    p_{DE}(z) = \frac{2n(1 + z)H(z)H'(z) - 6nH^2(z)}{\lambda\sqrt{\frac{H^{2}(z)}{\lambda H_{0}^{2}}}}.
\end{equation}
The EoS for DE can be determined which is
\begin{equation}\label{Eq:M2wde}
    \omega_{DE}(z) = - 1 + \frac{1}{3}(1 + z)\frac{H'(z)}{H(z)}.
\end{equation}
Using  equations (\ref{FE:18}), (\ref{FE:19}) and (\ref{FE:20}), we obtain the following
\begin{equation}\label{Eq:M2reff}
    \rho_{eff}(z) = 3H_{0}^{2}[\Omega_{m0}(1 + z)^{3} + \Omega_{r0}(1 + z)^{4}] + 6nH_{0}\sqrt{\frac{1}{\lambda}}H(z),
\end{equation}
\begin{equation}\label{Eq:M2peff}
    p_{eff}(z) = \frac{\lambda H_{0}^{2}\Omega_{r0}(1 + z)^{4}\sqrt{\frac{H^{2}(z)}{\lambda H_{0}^{2}}} + 2n(1 + z)H(z)H'(z) - 6nH^2(z)}{\lambda\sqrt{\frac{H^{2}(z)}{\lambda H_{0}^{2}}}},
\end{equation} 
\begin{equation}\label{Eq:M2weff}
    \omega_{eff}(z) = \frac{\lambda H_{0}^{2}\Omega_{r0}(1 + z)^{4}\sqrt{\frac{H^{2}(z)}{\lambda H_{0}^{2}}} + 2n(1 + z)H(z)H'(z) - 6nH^2(z)}{\lambda\sqrt{\frac{H^{2}(z)}{\lambda H_{0}^{2}}}\left(3H_{0}^{2}[\Omega_{m0}(1 + z)^{3} + \Omega_{r0}(1 + z)^{4}] + 6nH_{0}\sqrt{\frac{1}{\lambda}}H(z)\right)}.
\end{equation}
The parameter $\lambda$ can be expressed in terms of the other cosmological parameters using equations (\ref{FE:9}) and (\ref{M2}) which is given by
\begin{equation}\label{Eq:lambda}
    \lambda = \frac{4n^{2}}{(1 - \Omega_{m0} - \Omega_{r0})^{2}}.
\end{equation}

\subsection{Model-III}
Consider an exponential functional $F(Q)$ \cite{anagnostopoulos2021first}, given by
\begin{equation}\label{M3}
    F(Q) = Q \; e^{\beta \left(\frac{Q_{0}}{Q} \right)} - Q
\end{equation}
where $\beta$ is a constant, and $Q_{0} = 6H_{0}^{2}$ with $H_{0}$ representing present  Hubble parameter. The theory is equivalent to GR without a cosmological constant when $\beta = 0$. However, the exponential model fits observations satisfactorily   for a non zero $\beta $ and   absence of a cosmological constant in the $F(Q)$ function \cite{anagnostopoulos2021first} leads to a realistic scenario.  Here, the past asymptotic behaviour ($\frac{Q_{0}}{Q} \ll 1$) of the model can be  recovered by a pure GR at an early time and hence the model trivially pass the BBN constraints test \cite{anagnostopoulos2023new}.

Using equation (\ref{M3}), we estimate the DE density and the pressure , which are
\begin{equation}\label{Eq:M3rde}
    \rho_{DE}(z) = \frac{1}{2}\left(6H^{2}(z) - 6e^{\frac{\beta H^{2}_{0}}{H^{2}(z)}}H^{2}(z) + 12H^{2}_{0}\beta e^{\frac{\beta H^{2}_{0}}{H^{2}(z)}}\right),
\end{equation}
\[p_{DE}(z) = - (1+z)H(z)H'(z)\]
\begin{equation}\label{Eq:M3pde}
    \times\Big(- 2 + \frac{2e^{\frac{\beta H^{2}_{0}}{H^{2}(z)}}\left(36H^{2}(z) - 36H_{0}^{2}\beta H^{2}(z) + 72H_{0}^{4}\beta^{2}\right)}{36H^{4}(z)}\Big) - \rho_{DE}(z)
\end{equation}
where $(')$ denotes differentiation w.r.t. $z$. The effective energy density and pressure are determined from equations (\ref{FE:18}) and (\ref{FE:19}) using equation (\ref{M3}), which are
\[\rho_{eff}(z) = 3H_{0}^{2}\left[\Omega_{m0}(1 + z)^{3} + \Omega_{r0}(1 + z)^{4}\right]\]
\begin{equation}\label{Eq:M3reff}
     \hspace{2cm}+ 3H^{2}(z) - \left[3H^{2}(z) - 6\beta H^{2}_{0}\right]e^{\frac{\beta H^{2}_{0}}{H^{2}(z)}},
\end{equation}
\[p_{eff}(z) = H_{0}^{2}\Omega_{r0}(1 + z)^{4} - 3H^{2}(z) + 2(1+z)H(z)H_{z}(z) + \Big[\left(3H^{2}(z) - 6\beta H^{2}_{0}\right)\]
\begin{equation} \label{Eq:M3peff}   
     - 2(1 + z)H(z)H_{z}(z)\left(1 + \frac{2\beta^{2}H_{0}^{2} - \beta H_{0}^{2}H^{2}(z)}{H^{4}(z)}\right)\Big]e^{\frac{\beta H^{2}_{0}}{H^{2}(z)}}.
\end{equation}
The effective EoS ($\omega_{eff}$), the ratio of $p_{eff}$ and $\rho_{eff}$ are determined from equations (\ref{Eq:M3peff}) and (\ref{Eq:M3reff}), respectively.

The evolution of the dynamical quantities are analyzed with redshift parameter $z$. We plot the scale factor, deceleration parameter and EoS parametrized quantities. In the next section, we study variation of the Hubble parameter $H(z)$  assuming a constant jerk parameter.

\section{The Hubble parameter}
\label{sec:HP}
For a constant jerk parameter $j = \frac{1}{aH^{3}}\frac{d^3a}{dt^3}$, the scale factor can be determined \cite{Pradhan_2023}. Thus constraining jerk parameter yields a Hubble parameter which is a function of the redshift parameter ($z$) can be used to study the evolutionary behaviour of the universe. The deceleration parameter $q$ provides information of the transition from the decelerating to the accelerating phase. The deceleration parameter is related to the jerk parameter  which is
\begin{equation}\label{eq:2}
j(z) = q(z) + 2q(z) +  \frac{dq}{dz} \; (1 + z).
\end{equation}
Using the equation we analyz the cosmological model in different $f(Q)$-gravity.  For a constant $j(z) = j_{0}$ the deceleration parameter is given by 
\[
q(z) = \frac{1}{4}\Big(- 1 + \sqrt{8j_{0} + 1}\tanh\Big[\frac{1}{2}\Big(2\;ArcTanh\Big(\frac{4\sqrt{8j_{0} + 1}q_{0} + \sqrt{8j_{0} + 1}}{8j_{0} + 1}\Big)
\]
\begin{equation}\label{qz}
    + \sqrt{8j_{0} + 1}\log(1+z)\Big)\Big]\Big),
\end{equation}
where $q_{0} = q(z=0)$. The first derivative of the Hubble parameter w.r.t. $z$ can be expressed in terms of deceleration parameter which is given by
\begin{equation}
    \frac{dH}{dz} = \frac{1 + q(z)}{(1 + z)} \; H.
\end{equation}
On integration we determine the Hubble parameter which is
\[H(z) = H_{0}\exp{\int_{0}^{z}\frac{(1 + q(z))}{1 + z}dz}\]
\[\;\;\;= Exp\Big[\frac{1}{4}\Big(3\log(1 + z) + 2\log\Big[\cos\Big[ArcTan\Big[\frac{1 + 4d}{\sqrt{-1 - 8j_{0}}}\Big] \]
\begin{equation}\label{Eq:H}
    - \frac{1}{2}\sqrt{- 1 - 8j_{0}}\log(1 + z)\Big]\Big] + \log\Big[\frac{8(j_{0} - d - 2d^{2})}{1 + 8j_{0}}\Big]\Big)\Big]H_{0},    
\end{equation}
where $H_{0}$ is the present value. The late accelerating phase of the universe for three different functional forms of $f(Q)$ gravity are analyzed making use of observational datasets and estimating $H_{0}$, $j_{0}$ and $q_{0}$ in the next section.

\section{Observational constraints on $H_{0}$, $j_{0}$ and $q_{0}$}
\label{sec: obs}
In this section, the observational Hubble data and Pantheon data are used to estimate contraints $H_{0}$, $j_{0}$ and $q_{0}$ here.
\subsection{For Hubble Dataset}
The Hubble parameter $H(z)$ is measured from the line-of-sight of BAO data which includes the correlation functions of the luminous red galaxies and the differential age (DA) of the galaxies. Recently, a list of 57 data points was compiled by Sharov and Vasiliev for $H(z)$ in the redshift range $0.07 \le z \le 2.42$ \cite{Sharov_2018}. The dataset includes 31 points measured by the DA method (also known as the cosmic chronometer (CC) technique) and 26 from BAO and other measurements. Here we use 31 data points of $H(z)$ as shown in Table \ref{tabd} in the redshift interval $0.07 \le z \le 1.965$ to estimate the best-fit values of the parameters making use of {\it chi-square} minimization technique. Since the data points of $H(z)$ are uncorrelated, we define $\chi^{2}_{CC}$, which is
\begin{equation}\label{chi_cc}
    \chi^{2}_{CC}(H_{0}, j_{0}, q_{0}) = \sum_{i=1}^{57}\frac{\left[H_{th}(z_{i}, H_{0}, j_{0}, q_{0}) - H_{obs}(z_{i})\right]^{2}}{\sigma_{H}^{2}(z_{i})},
\end{equation}
where $H_{th}$ is the theoretical value of the Hubble parameter estimated in equation (\ref{Eq:H}), $H_{obs}(z_{i})$ is the observed Hubble parameter in redshift $z_{i}$ and $\sigma_{H}(z_{i})$ is the standard deviation.
\begin{table*}
\centering
\caption{$H(z)-z$ dataset with standard errors using in the current analysis (the Hubble parameters in km s$^{- 1}$Mpc$^{- 1}$)}
\begin{tabular}{cccccccc}\hline
&       &    Hubble data  &       &\\
\hline
$z$ & $H(z)$ & $\sigma_{H}$ & Ref. & $z$ & $H(z)$ & $\sigma_{H}$ & Ref.\\
\hline
0.07 & 69 & 19.6 & \cite{Cong_2014h} & 0.4783 & 80.9 & 9 & \cite{Moresco_2016}\\
0.09 & 69 & 12 & \cite{Daniel_Stern_2010} & 0.480 & 97 & 62 & \cite{Daniel_Stern_2010}\\
0.120 & 68.6  & 26.2 & \cite{Cong_2014h} & 0.593 & 104 & 13 & \cite{M_Moresco_2012}\\
0.170 & 83 & 8 & \cite{Daniel_Stern_2010} & 0.6797 & 92 & 8 & \cite{M_Moresco_2012} \\
0.1791 & 75 & 4 & \cite{M_Moresco_2012} & 0.7812 & 105 & 12 & \cite{M_Moresco_2012}\\
0.1993 & 75 & 5 & \cite{M_Moresco_2012} & 0.8754 & 125 & 17 & \cite{M_Moresco_2012}\\
0.200 & 72.9 & 29.6 & \cite{Cong_2014h} & 0.880 & 90 & 40 & \cite{Daniel_Stern_2010} \\
0.270 & 77 & 14  & \cite{Daniel_Stern_2010} & 0.900 & 117 & 23 & \cite{Daniel_Stern_2010}\\
0.280 & 88.8 & 36.6 & \cite{Cong_2014h} & 1.037 & 154 & 20 & \cite{M_Moresco_2012}\\
0.3519 & 83 & 14 & \cite{M_Moresco_2012} & 1.300 & 168 & 17 & \cite{Daniel_Stern_2010}\\
0.3802 & 83 & 13.5 & \cite{Moresco_2016} & 1.363 & 160 & 33.6 & \cite{Moresco_2015}\\
0.400 & 95 & 17 & \cite{Daniel_Stern_2010} & 1.430 & 177 & 18 & \cite{Daniel_Stern_2010}\\
0.4004 & 77 & 10.2 & \cite{Moresco_2016} & 1.530 & 140 & 14 & \cite{Daniel_Stern_2010}\\
0.4247 & 87.1 & 11.2 & \cite{Moresco_2016} & 1.750 & 202 & 40 & \cite{Daniel_Stern_2010}\\
0.4497 & 92.8 & 12.9 & \cite{Moresco_2016} & 1.965 & 186.5 & 50.4 & \cite{Moresco_2015}\\
0.470 & 89 & 34 & \cite{Ratsimbazafy_2017} &  &   &  \\
\hline
\end{tabular}
\label{tabd}
\end{table*}

\subsection{Observational Analysis with Pantheon Dataset}
\label{}
Here the observed Pantheon supernovae type Ia (SNe Ia) dataset taken into account to constrain the model parameters which consists of  spectroscopically confirmed 1048 supernovae specimens compiled data by Scolnic et al. \cite{Scolnic_2018}, the sample consists of different supernovae surveys both at the high and the low redshift regimes, namely, the CfA1-CfA4 surveys \cite{Hicken_2009}, the PanSTARRS1 (PS1) survey \cite{Scolnic_2018}, Sloan Digital Sky Survey (SDSS) \cite{Sako_2018}, SuperNovae Legacy Survey (SNLS) \cite{Guy_2010}, ESSENCE \cite{Narayan_2016}, Carnegie Supernova project (CSP) \cite{Contreras_2010} and other Hubble space telescope (HST) data \cite{Graur_2014}, \cite{Riess_2018}, \cite{Riess_2007}. The range of redshift for Pantheon sample lies $0.01 < z < 2.26$. (For detail please see Ref. \cite{Asvesta_2022}). We define the $\chi^{2}_{SN}$ function from the Pantheon sample of 1048 SNe Ia as given by
\begin{equation}\label{eq26}
\chi^{2}_{SN}(H_{0}, j_{0}, q_{0}) = \Delta \mathcal{F}_{i} C_{SN}^{-1}\Delta \mathcal{F}_{j},  
\end{equation}
where $\Delta \mathcal{F} = \mathcal{F}_{th} - \mathcal{F}_{obs}$ is the difference between the theoretical and the observed values of the apparent magnitude for each SNe Ia at s redshift $z_{i}$, and $C_{SN}$ is the total covariance matrix \cite{Scolnic_2018}. For SNe Ia  we take distance modulus formula:
\begin{equation}\label{eq22}
    m(z) = M + 5\;\log_{10}\Big[\frac{d_{L}(z)}{1Mpc}\Big] + 25,
\end{equation}
where $m$, $M$ are the apparent and absolute magnitudes and $d_{L}(z)$ is the luminosity distance for a flat universe which is 
\begin{equation}\label{eq23}
    d_{L}(z) = c(1 + z)\int_{0}^{z}\frac{dz'}{H(z')},
\end{equation}
with redshift $z$ for SNe Ia in the CMB rest frame. Define Hubble free luminosity distance as $D_{L}(z)\equiv \frac{H_{0}d_{L}(z)}{c}$, we get theoretical apparent magnitude which is 
\begin{equation}\label{eq24}
    m(z) = M + 5\log_{10}[D_{L}(z)] + 5\log_{10}\Big(\frac{c/H_{0}}{Mpc}\Big) + 25. 
\end{equation}
However, a degeneracy between $M$ and $H_{0}$ are observed, thus a new parameter $\mathcal{M}$ can be defined as,
\begin{equation}\label{eq25}
    \mathcal{M} = M + 5\log_{10}\Big[\frac{c/H_{0}}{1Mpc}\Big] + 25
\end{equation}
A number of attempts have been made to marginalize the degenerate combination and recently in \cite{Asvesta_2022} minimized the parameter using the Pantheon sample for a tilted universe and  $\mathcal{M}$ is close to 23.8. 

\begin{figure}[t]
    \centering
    \includegraphics[width=0.8\textwidth]{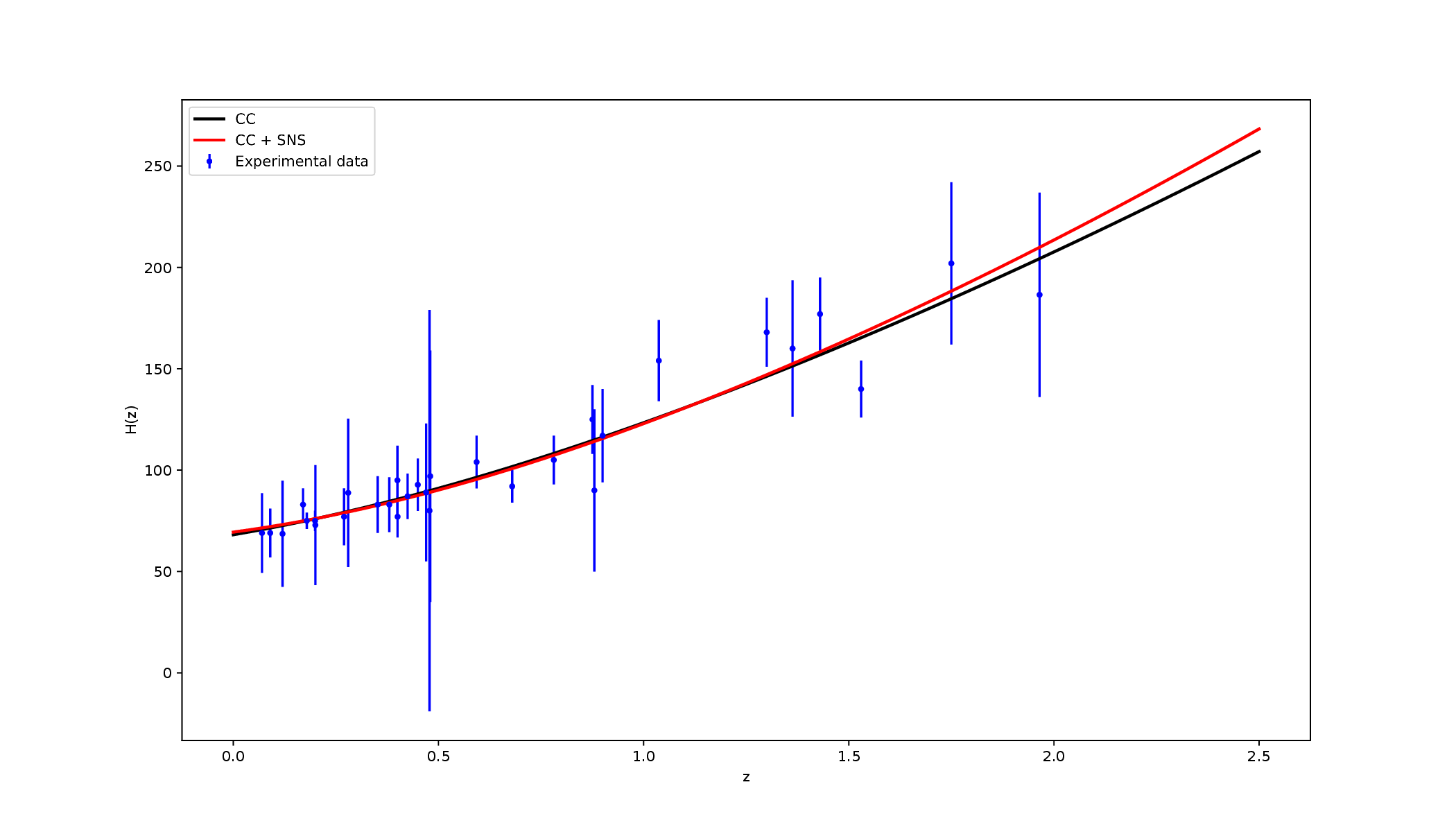}
    \caption{Variation of $H(z)$ with respect to redshift parameter ($z$). The black curve for best-fit values $H_{0} = 68.13$, $j_{0} = 0.93$ and $q_{0} = - 0.45$. The red curve for best-fit values $H_{0} = 69.418$, $j_{0} = 1.208$ and $q_{0} = - 0.604$. The dots corresponds to the Hubble data with the error bar.}
    \label{fit:HD}       
\end{figure}

\begin{figure}[!]
    \includegraphics[width=0.8\textwidth]{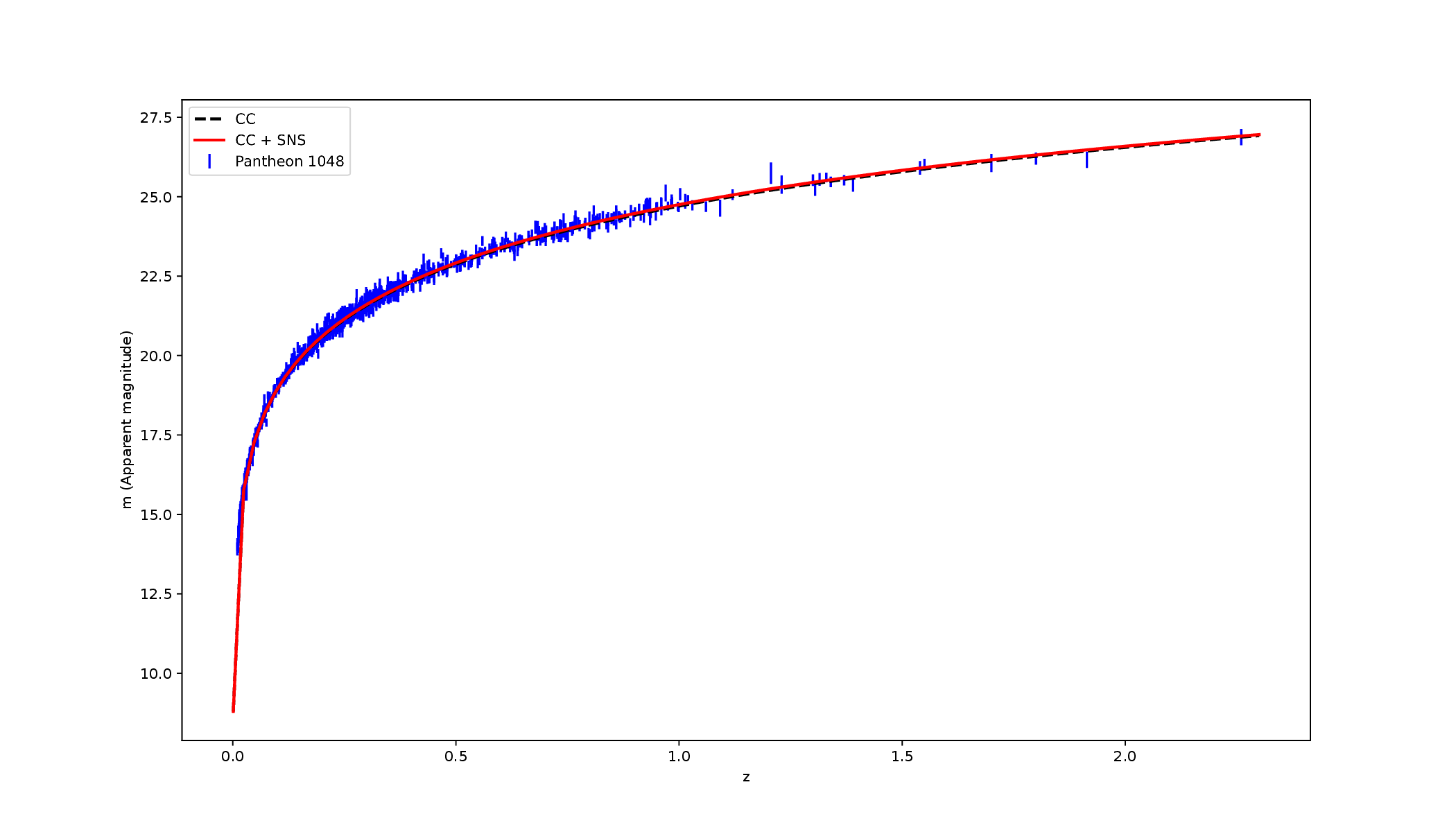}
    \caption{Variation of apparent magnitude $m(z)$ with respect to $z$. The black dashed curve for best-fit values $H_{0} = 68.13$, $j_{0} = 0.93$ and $q_{0} = - 0.45$. The red curve for best-fit values $H_{0} = 69.418$, $j_{0} = 1.208$ and $q_{0} = - 0.604$. The dots corresponds to the 1048 Pantheon dataset with the error bar}
    \label{fit:SN} 
\end{figure}

We use here two different analyses for constraining the parameters $H_{0}$, $j_{0}$ and $q_{0}$ using cosmic chronometers (CC) and jointly CC and Pantheon datasets, respectively. For joint analysis, we use $\chi_{total}^{2}$ as:
\begin{equation}\label{chi_total}
    \chi^{2}_{total} = \chi^{2}_{CC} + \chi^{2}_{SN}.
\end{equation}
The best-fit values are obtained by minimizing the {\it chi-square} functions given in equations (\ref{chi_cc}) and (\ref{chi_total}). The best-fit values of the parameters $H_{0}$, $j_{0}$ and $q_{0}$ are tabulated in Table \ref{table:2}. Thereafter, best-fit curves for the Hubble datasets and Pantheon datasets with error bars are drwan in Figs. \ref{fit:HD} and \ref{fit:SN}. The Hubble constant was estimated by the Planck Collaboration in 2018 \cite{planck_2018_results} to be $H_{0} = 67.4 \pm 0.5$ km s$^{- 1}$Mpc$^{- 1}$, whereas in 2021 Riess {\it et al.} \cite{Riess_2021} predicted the value $H_{0} = 73.2 \pm 1.3$ km s$^{- 1}$Mpc$^{- 1}$. Other observational groups recently estimated and obtained Hubble parameter different from the value. Cao and Ratra \cite{Cao_2023} predicted the Hubble constant $H_{0} = 69.8 \pm 1.3$ km s$^{- 1}$Mpc$^{- 1}$, whereas $H_{0} = 69.7 \pm 1.2$ km s$^{- 1}$Mpc$^{- 1}$ was the estimate made in \cite{Cao_2022}. In the likelihood analysis of extensive observational datasets, Alberto Domnnguez {\it et al.} \cite{Dom_nguez_2019} obtained the  Hubble parameter which is $H_{0} = 66.6 \pm 1.6$ km s$^{- 1}$Mpc$^{- 1}$, while \cite{Park_2020,Lin_2021} determined $H_{0} = 65.8 \pm 3.4$ km s$^{- 1}$Mpc$^{- 1}$. Freedman {\it et al.} \cite{Freedman_2020} estimated the present value of Hubble constant $H_{0} = 69.6 \pm 0.8$ km s$^{- 1}$Mpc$^{- 1}$, Birrer {\it et al.} \cite{Birrer_2020} measured $H_{0} = 67.4^{+ 4.1}_{- 3.2}$ km s$^{- 1}$Mpc$^{- 1}$, Boruah {\it et al.} \cite{Boruah_2021}  measured $H_{0} = 69^{+ 2.9}_{- 2.8}$ km s$^{- 1}$Mpc$^{- 1}$ and most recently, Freedman \cite{Freedman_2021} gave an estimated value $H_{0} = 69.8 \pm 0.6$ km s$^{- 1}$Mpc$^{- 1}$ and Qin Wu {\it et al.} \cite{Wu_2022}  estimated $H_{0} = 68.81^{+ 4.99}_{- 4.33}$ km s$^{- 1}$Mpc$^{- 1}$.
%

\begin{table}[h]
\caption{Best fit values of $H_{0}$, $j$ and $q_{0}$}
\label{table:2}       
\begin{tabular}{|c|c|c|c|c|}
\hline
Datasets & $H_{0}$ & $j_{0}$ & $q_{0}$ & $\chi_{min}^{2}$  \\
\hline
CC & 68.13 & 0.93 & -0.45 & 13.602 \\
\hline
CC + Pantheon & 69.418 & 1.208 & -0.604 & 1040.583\\
\hline
\end{tabular}
\end{table}

\section{Analysis of cosmological parameters}
\label{sec:evo}

\subsection{Analysis of deceleration parameter}
\begin{figure*}[t]
    \centering
    \includegraphics[width=0.8\textwidth]{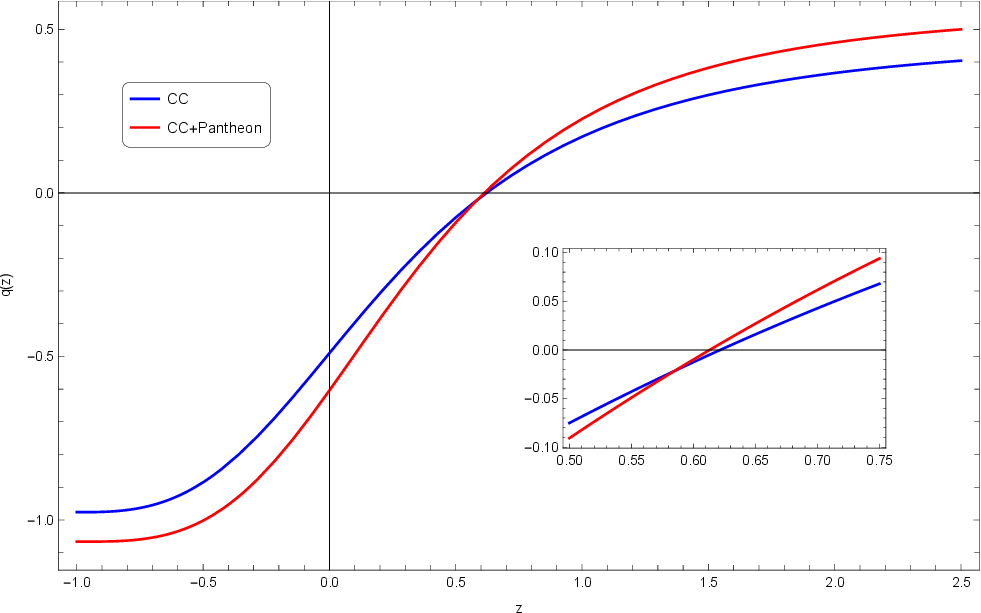}
    \caption{Evolution of the deceleration parameter $q(z)$ with the redshift parameter ($z$). The blue and red curves for the best-fit values of $H_{0}$, $j_{0}$, and $q_{0}$ from the study of the CC and combined CC + Pantheon data, respectively.}
    \label{fig:dece_ccp}       
\end{figure*}

We analyze deceleration parameter $q(z)$ using equation (\ref{qz}). The deceleration parameter is plotted as a function of the redshift parameter ($z$) in Fig. \ref{fig:dece_ccp}. It is evident and it is passing through the deceleration phase flip sign at a recent past. We note that a deceleration to acceleration change over at a redshift $z_{t} = 0.62$ and $z_{t} = 0.61$ for the best-fit values obtained from CC and joint analysis of CC+Pantheon datasets. 

\begin{figure}[t]
    \centering
    \includegraphics[width=0.8\textwidth]{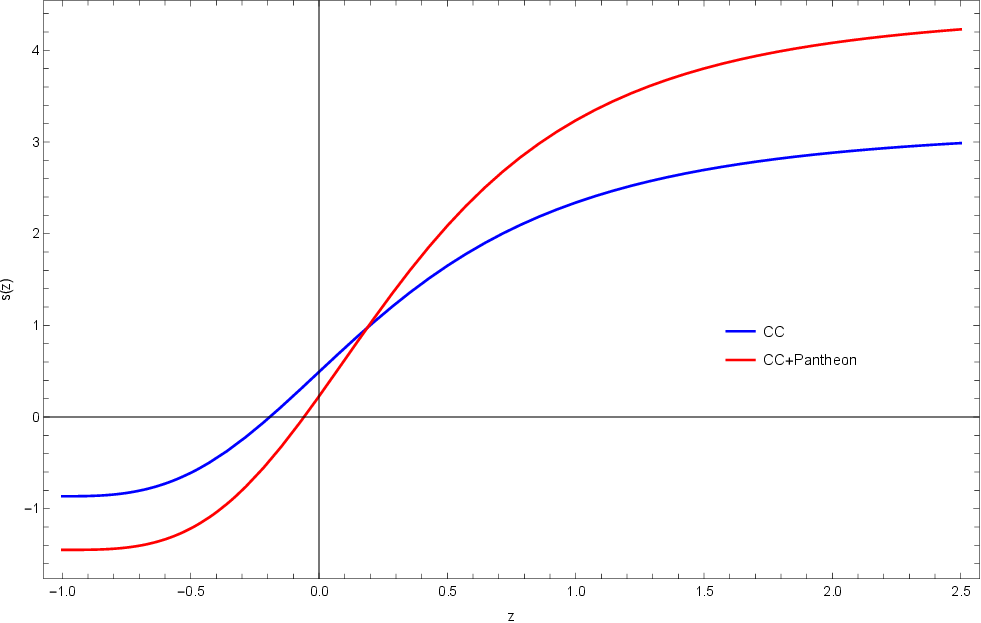}
    \caption{Evolution of the Snap parameter $``s(z)"$ with the redshift parameter ($z$). The blue and red curves for the best-fit values of $H_{0}$, $j_{0}$ and $q_{0}$ from the study of the CC and combined CC + Pantheon data, respectively.}
    \label{fig:snap_ccp}       
\end{figure}

\subsection{Snap parameter}
Here we investigate one of the  state finder diagnostic parameter related to late time evolution of the universe namely, snap parameter $s(z)$ \cite{Sahni_2003} which is given by
\begin{equation}\label{eq:s}
    s(z) =  (3q(z) + 2)j(z) + (1 + z)\frac{dj(z)}{dz}.
\end{equation}
For $j(z) = j_{0}$, the snap parameter becomes
\begin{equation}\label{eq:ss}
    s(z) =  (3q(z) + 2)j_{0}.
\end{equation}
We plot the snap parameter $s(z)$ in Fig. \ref{fig:snap_ccp}. It is evident that the snap parameter lies in the range $(0,1)$ at the present epoch.

\subsection{Energy density and EoS parameters}
\label{sec:edseos}
In section \ref{cosmo:models}, energy density and EoS parameter for DE and the effective fluids are determined for the three models: power-law, log-square root, and exponential functional of $f(Q)$ gravity. The parameters for the three models were chosen in such a way so as to ensure that the energy density remains positive. The evolution of EoS parameters helps us to understand how the universe transits through various stages over time, offering insights into the behaviour of DE and its impact on cosmic expansion.
\subsubsection{Power-law {\ensuremath{f(Q)}} model}
\begin{figure}[t]
\centering
\begin{subfigure}{0.5\textwidth}
    \centering
    \includegraphics[width=6cm]{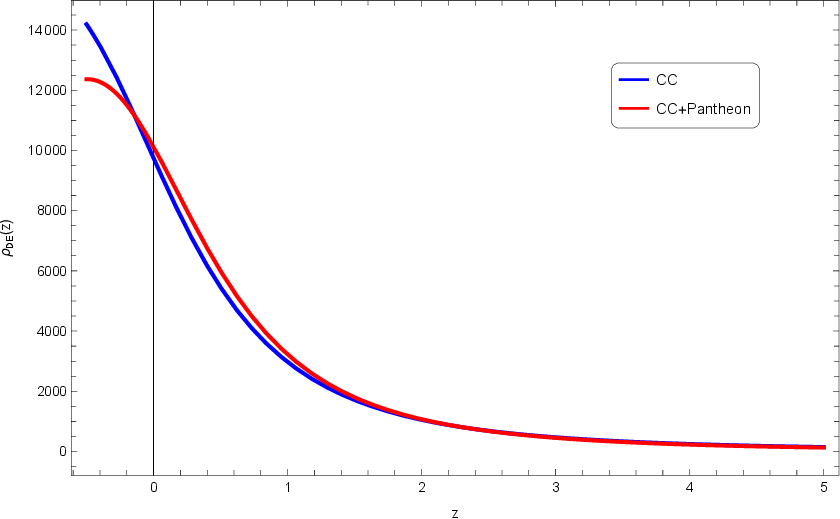}
    \caption{}
    \label{M1red}       
\end{subfigure}%
\begin{subfigure}{0.5\textwidth}
    \centering
    \includegraphics[width=6cm]{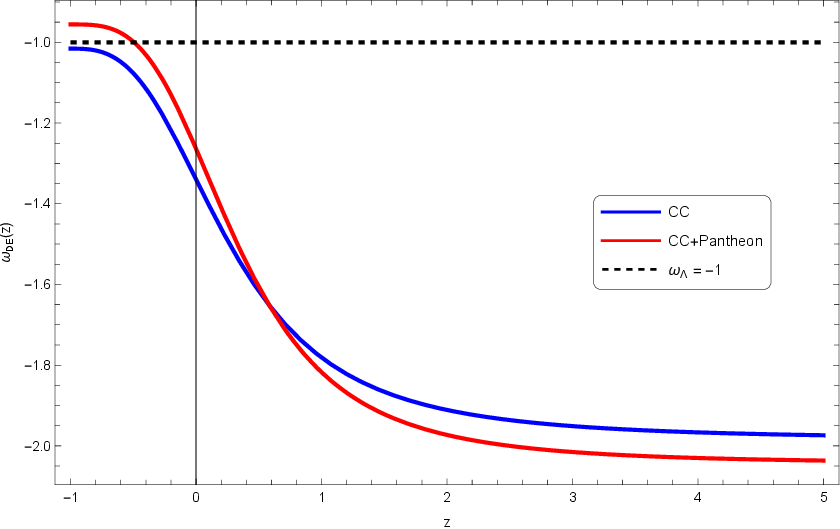}
    \caption{}
    \label{M1wed}
\end{subfigure}
\caption{DE density $\rho_{DE}(z)$ (Left panel) and EoS parameter $\omega_{DE}(z)$ (Right panel)  with redshift parameter ($z$) for the power law model with $n = -1$. The blue and red curves for the best fit values of $H_{0}$, $j_{0}$ and $q_{0}$ from the study of the CC and combined CC + Pantheon data, respectively.}
\label{M1rw}       
\end{figure}

\begin{figure}[h]
\centering
\begin{subfigure}{0.5\textwidth}
    \centering
    \includegraphics[width=6cm]{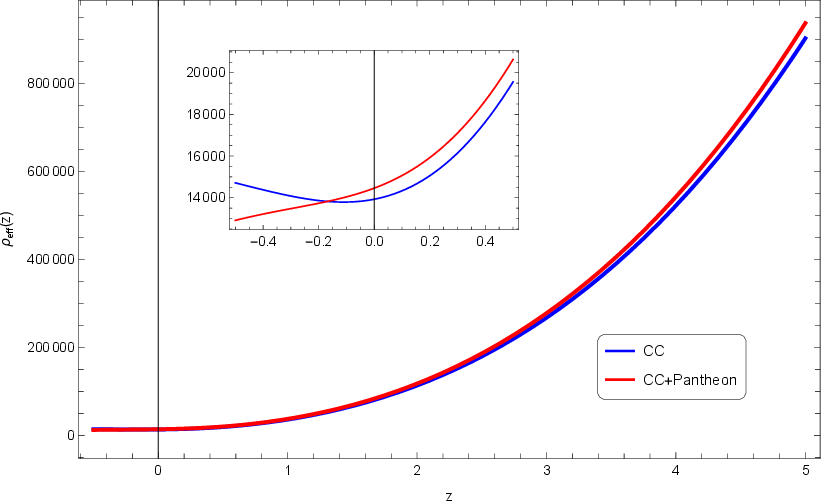}
    \caption{}
    \label{M1reff}       
\end{subfigure}%
\begin{subfigure}{0.5\textwidth}
    \centering
    \includegraphics[width=6cm]{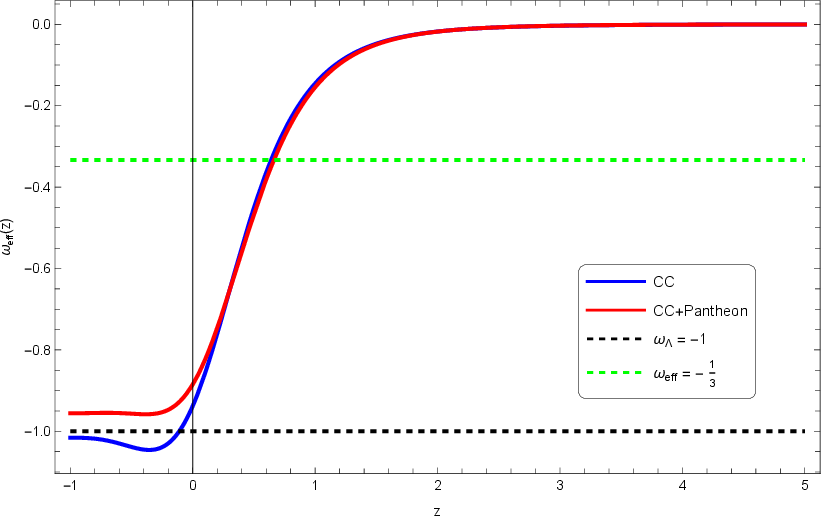}
    \caption{}
    \label{M1weff}
\end{subfigure}
\caption{The evolution of effective energy density $\rho_{eff}(z)$ (Left panel) and EoS parameter $\omega_{eff}(z)$ (Right panel) as a function of the redshift parameter ($z$) for the power law model with model parameters $n = -1$. The blue and red curves for the best-fit values of CC and CC + Pantheon datasets, respectively.}
\label{M1rweff}       
\end{figure}

\begin{figure}
\centering
\begin{subfigure}{0.5\textwidth}
    \centering
    \includegraphics[width=6cm]{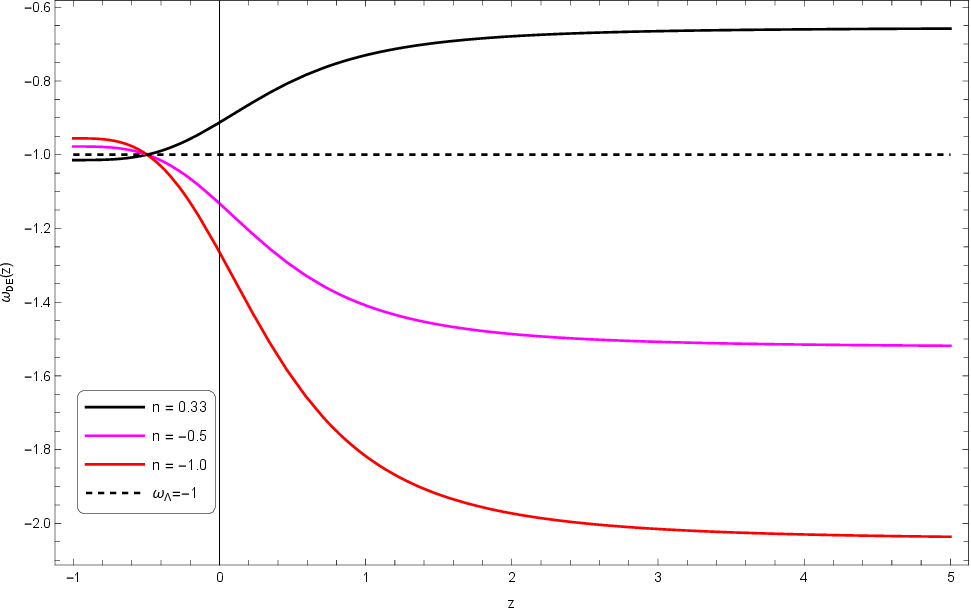}
    \caption{}
    \label{m1wden}       
\end{subfigure}%
\begin{subfigure}{0.5\textwidth}
    \centering
    \includegraphics[width=6cm]{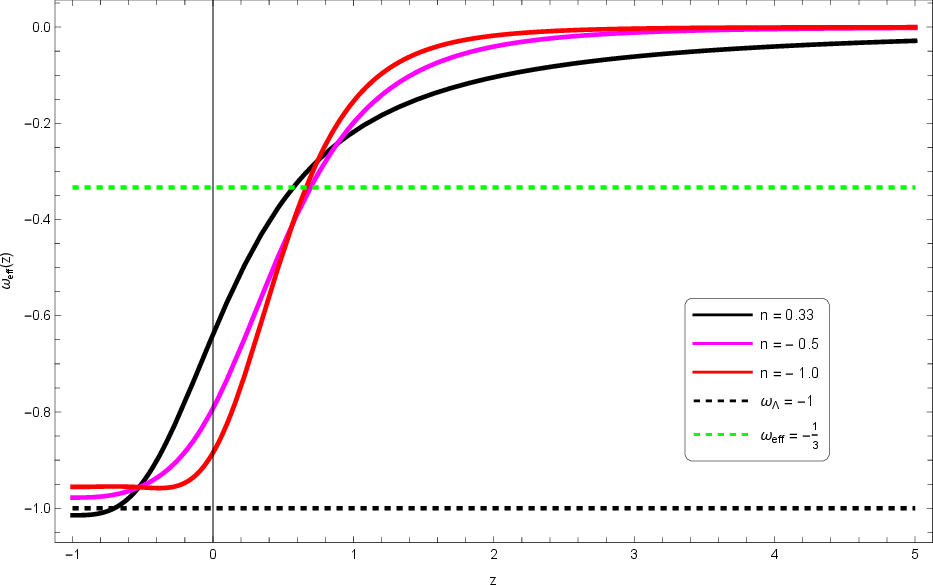}
    \caption{}
    \label{m1weffn}
\end{subfigure}
\caption{The variation of EoS parameter $\omega_{DE}(z)$ and $\omega_{eff}(z)$ as a function of the redshift parameter ($z$) for the power law model with model parameters $n = 0.33$, $-0.5$, and $-1$.}
\label{M1wdeffn}       
\end{figure}

In Fig. \ref{M1rw}, we draw evolution of the DE density, $\rho_{DE}(z)$ (left panel) and  $\omega_{DE}(z)$ (right panel) with redshift $z$ using the best-fit values of $H_{0}$, $j_{0}$ and $q_{0}$ provided in Table \ref{table:2}. We consider here the model parameter $n = -1$. From the left panel, it is evident that the DE density $\rho_{DE}(z)$ increases as the universe expands, indicating a more positive energy density over time. The right panel reveals that the DE EoS parameter $\omega_{DE}$ converges to $-1$ as the redshift decreases, highlighting a trend towards a cosmological constant-like behaviour in future. At the present epoch ($z=0$), the EoS parameter is found to hover in the phantom region, with $\omega_{DE} < - 1$. The evolutionary behaviour of effective energy density, $\rho_{eff}(z)$ and effective EoS, $\omega_{eff}(z)$ are plotted Fig. \ref{M1rweff}. The effective energy density is found positive, which indicates a consistent contribution to the overall energy content of the universe. The effective EoS parameter, on the other hand is decreasing  during early times which later is found to become negative. A decelerating phase characterized by $\omega_{eff} > - \frac{1}{3}$ is found, which then transits  to an accelerating phase at a redshift, where $\omega_{eff} = - \frac{1}{3}$. At the present epoch, the effective EoS parameter  $\omega_{eff} = - 0.89$ based on the best-fit values from cosmic chronometers (CC) data and $\omega_{eff} = - 0.94$ from the combined CC and Pantheon datasets. In the future, the effective EoS parameter will attain  $-1$ through quintessence like evolution. A transition from quintessence to phantom for best-fit values of CC and CC + pantheon, respectively are predicted in the models. The effective EoS parameter: $\omega_{eff} = - 1$, indicating a future dominated by a cosmological constant-like DE component. Further, for different values of the model parameter $n$, the variation of the EoS parameter for DE and effective fluid of the universe is plotted in Fig. \ref{M1wdeffn}. For $n = 0.33$, $\omega_{DE}$ decreases as universe expand, where as increases for the values of $n = - 0.5, -1$ and the value approaches to $- 1$ in the future. The DE behaves like quintessence for $n > 0$ and phantom for $n < 0$, at a late universe, whereas in the future, it will behave as a cosmological constant. But, $\omega_{eff}$ is found to decrease for $n = 0.33, - 0.5, - 1$ from early to present time.

\subsubsection{Log-square root model}
\begin{figure}[h]
\centering
\begin{subfigure}{0.5\textwidth}
    \centering
    \includegraphics[width=6cm]{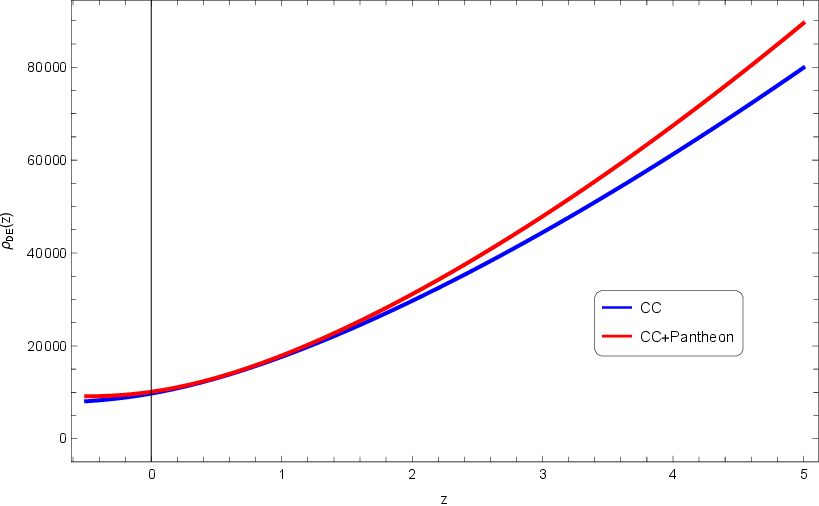}
    \caption{}
    \label{M2rde}       
\end{subfigure}%
\begin{subfigure}{0.5\textwidth}
    \centering
    \includegraphics[width=6cm]{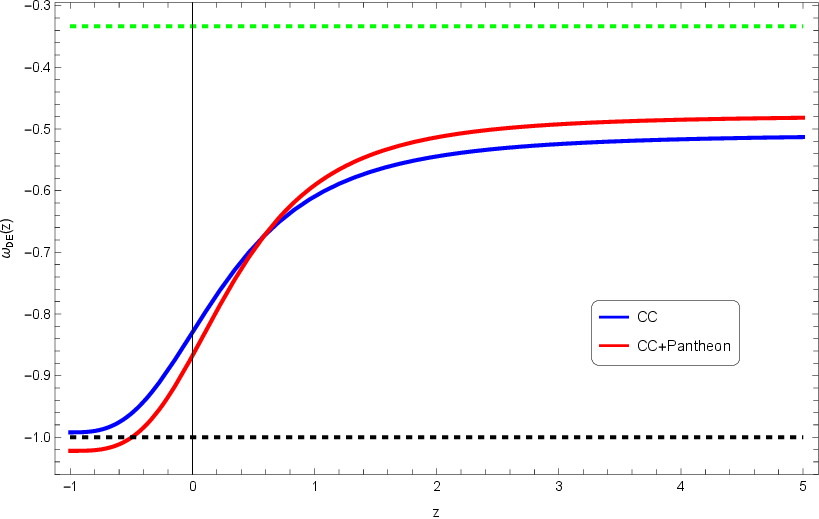}
    \caption{}
    \label{M2wde}
\end{subfigure}
\caption{The evoluation of $\rho_{DE}(z)$ and EoS parameter $\omega_{DE}(z)$ as a function of the redshift parameter ($z$) for the Log-square-root model. The blue and red curves correspond to the best-fit values of CC and combined CC + Pantheon, respectively.}
\label{M2rwde}       
\end{figure}

In the log-square root model, the DE density, $\rho_{DE}(z)$, (left pane) and the EoS parameter, $\omega_{DE}(z)$, (right panel) are ploted as a function of redshift parameter ($z$) in Fig. \ref{M2rwde} using the best-fit values displayed in Table \ref{table:2}. It is evident that the DE energy density decreases as redshift parameter decreases but it remains  positive throughout the evolution. The EoS parameter, $\omega_{DE}(z)$, exhibits quintessence behaviour, characterized by $- 1 < \omega_{DE} < - \frac{1}{3}$, from the early universe to the present time. The CC data predicts  $\omega_{DE} \geq -1$ which signifies DE initially acts as a dynamical field (quintessence) but eventually, it behaves like a cosmological constant at a late time, but the CC+ Pantheon data predicts $\omega_{DE} < -1$ which implies a universe with phantom dominated matter.

\begin{figure}[h]
\centering
\begin{subfigure}{0.5\textwidth}
    \centering
    \includegraphics[width=6cm]{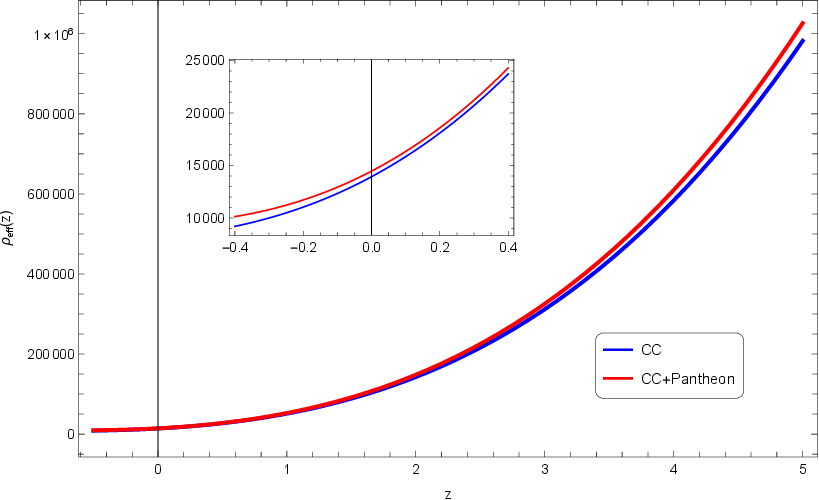}
    \caption{}
    \label{M2reff}       
\end{subfigure}%
\begin{subfigure}{0.5\textwidth}
    \centering
    \includegraphics[width=6cm]{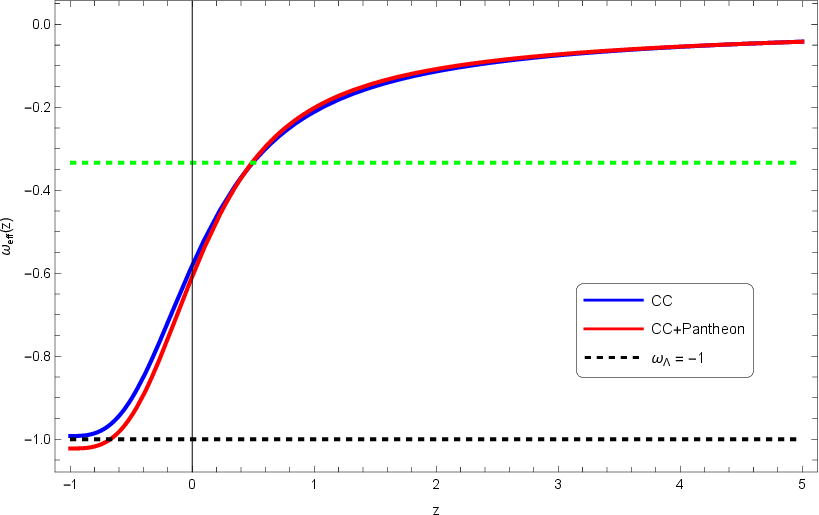}
    \caption{}
    \label{M2weff}
\end{subfigure}
\caption{The evoluation of $\rho_{eff}(z)$ and EoS parameter $\omega_{eff}(z)$ as a function of the redshift parameter ($z$) for the Log-square-root model. The blue and red curves correspond to the best-fit values of CC and combined CC + Pantheon, respectively.}
\label{M2rweff}       
\end{figure}

The behaviour of effective energy density, $\rho_{eff}(z)$, in the left panel and effective EoS, $\omega_{eff}(z)$, in the right panel drawn in Fig. \ref{M2rweff} indicates that the effective energy density $\rho_{eff}(z)$ decreases as the universe expands which remains positive throughout its evolution. The behaviour of effective EoS parameter signifies that the universe began from a decelerating phase, characterized by $\omega_{eff} > - \frac{1}{3}$ later transits to an accelerating phase with  $\omega_{eff} < - \frac{1}{3}$. The EoS parameter is found to converge to $\omega_{eff} = -1$ in future, indicating the log-square root model accommodates a universe a cosmological constant. At the present epoch, the effective EoS parameter is  $\omega_{eff} = - 0.6$ and $\omega_{eff} = - 0.6$ obtained for the best-fit values estimated using  cosmic chronometers (CC) data and the combined CC + Pantheon datasets.

\subsubsection{Exponential model}
\begin{figure}[t]
\centering
\begin{subfigure}{0.5\textwidth}
    \centering
    \includegraphics[width=6cm]{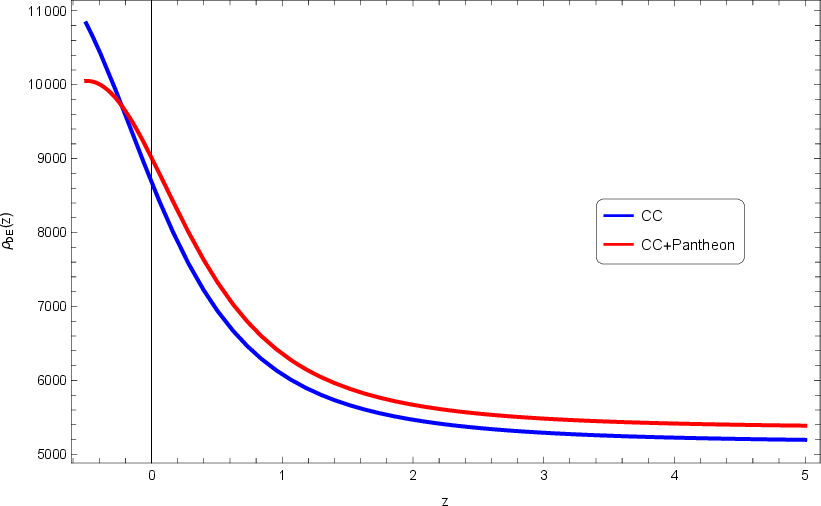}
    \caption{}
    \label{M3rde}       
\end{subfigure}%
\begin{subfigure}{0.5\textwidth}
    \centering
    \includegraphics[width=6cm]{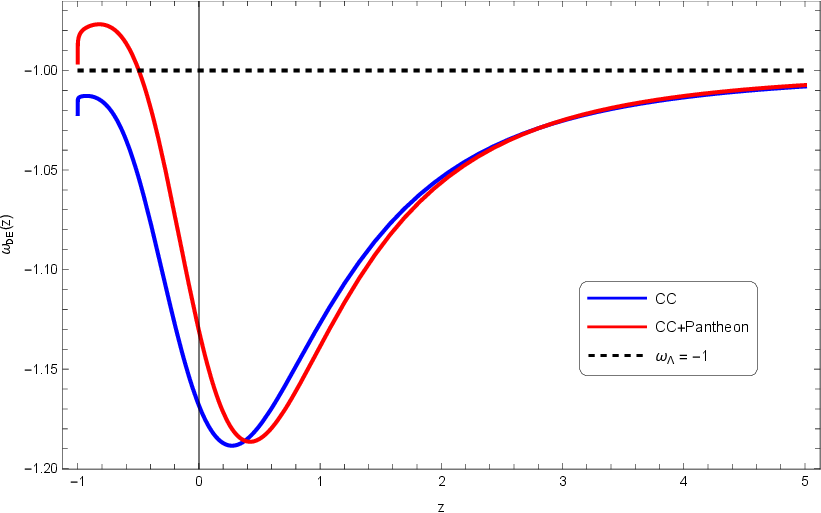}
    \caption{}
    \label{M3wde}
\end{subfigure}
\caption{The evolution of $\rho_{DE}(z)$ (left panel) and EoS parameter $\omega_{DE}(z)$ (right panel) as a function of the redshift $z$ for the exponential model with $\beta = 0.37$. The blue and red curves correspond to the best-fit values of CC and combined CC + Pantheon, respectively.}
\label{M3rwde}       
\end{figure}

In this case, the DE density $\rho_{DE}(z)$ and EoS parameter $\omega_{DE}(z)$ are plotted in Fig. \ref{M3rwde} for $\beta = 0.37$ and best-fit values taken from Table \ref{table:2}. In the left panel of the figure, it is evident that DE density increases as the universe expands. We note that the observed EoS parameter $\omega_{DE}$ to begin with was $< -1$ in both CC and CC+ Pantheon data prediction which however decreases, attains a minimum value there after increases. In future, we note that $\omega_{DE} < -1$ for former case and $\omega_{DE}>-1$ for later case respectively. Thus although a phantom regime exist from  early to future time for CC data, a different DE noted for CC+Pantheon data that permit a transition from phantom to quintessence in future.  

\begin{figure}[h]
\centering
\begin{subfigure}{0.5\textwidth}
    \centering
    \includegraphics[width=6cm]{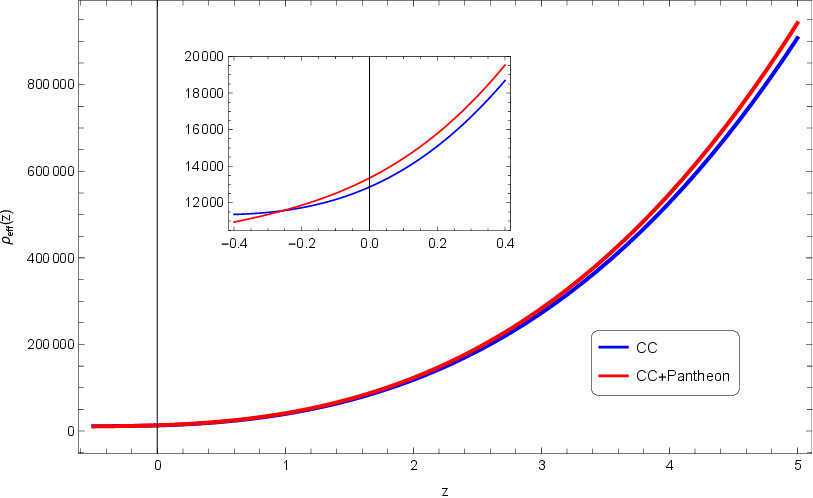}
    \caption{}
    \label{M3reff}       
\end{subfigure}%
\begin{subfigure}{0.5\textwidth}
    \centering
    \includegraphics[width=6cm]{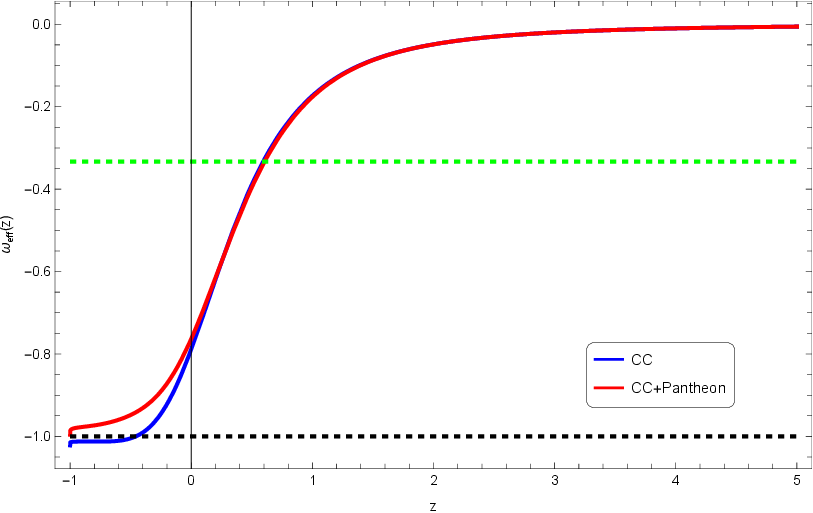}
    \caption{}
    \label{M3weff}
\end{subfigure}
\caption{The evolution of $\rho_{eff}(z)$ (left panel) and EoS parameter $\omega_{eff}(z)$ (right panel) as a function of the redshift $z$ for the exponential model with $\beta = 0.37$. The blue and red curves correspond to the best-fit values of CC and combined CC + Pantheon, respectively }
\label{M3rweff}       
\end{figure}

The behaviour of effective energy density $\rho_{eff}(z)$ and effective EoS parameter $\omega_{eff}(z)$ are plotted in Fig. \ref{M3rweff}. It is found that $\rho_{eff}(z)$ remain positive throughout the evolution of the universe, whereas the effective EoS parameter decreases from a large value.  The present values of the effective EoS parameter are $\omega_{eff} = - 0.79$ and $\omega_{eff} = - 0.76$ for the best-fit values of CC and CC + Pantheon datasets, respectively, which converge to $-1$ in the future.

\section{Energy conditions}
\label{sec:ECs}
The energy conditions (ECs) are crucial for an effective investigation of the cosmological geodesics. The well-known Raychaudhury equations can be used to obtain such conditions, which is given by
\begin{equation}\label{EC1}
    \frac{d\theta}{d\tau} = - \frac{1}{3}\theta^{2} - \sigma_{\mu\nu}\sigma^{\mu\nu} + \omega_{\mu\nu}\omega^{\mu\nu} - R_{\mu\nu}u^{\mu}u^{\nu},
\end{equation}
\begin{equation}\label{EC2}
    \frac{d\theta}{d\tau} = - \frac{1}{2}\theta^{2} - \sigma_{\mu\nu}\sigma^{\mu\nu} + \omega_{\mu\nu}\omega^{\mu\nu} - R_{\mu\nu}n^{\mu}n^{\nu},
\end{equation}
where $\theta$ is the expansion factor, $\sigma^{\mu\nu}$ and $\omega_{\mu\nu}$ are the shear and the rotation associated with the vector field $u^{\mu}$ and $n^{\mu}$ is the null vector \cite{PhysRev.98.1123,NOJIRI_2007}. For attractive gravity, equations (\ref{EC1}) and (\ref{EC2}) satisfy the following conditions:
\begin{equation}
    R_{\mu\nu}u^{\mu}u^{\nu} \ge 0,
\end{equation}
\begin{equation}
    R_{\mu\nu}n^{\mu}n^{\nu} \ge 0.
\end{equation}
Therefore, if we are working with a perfect fluid matter distribution, the energy conditions for $f(Q)$ gravity are given by 
\begin{itemize}
    \item[$\bullet$] {\bf Weak energy conditions(WEC):} 
    \begin{equation}\label{Eq:wec}
        \rho_{eff} \ge 0,\;\hspace{0.5cm}\; \rho_{eff} + p_{eff} \ge 0
    \end{equation}
    \item[$\bullet$] {\bf Null energy condition (NEC):} 
    \begin{equation}\label{Eq:nec}
        \rho_{eff} + p_{eff} \ge 0
    \end{equation}
    \item[$\bullet$] {\bf Dominant energy condition (DEC):}
    \begin{equation}\label{Eq:dec}
        \rho_{eff} \ge 0,\;\hspace{0.5cm}\; \rho_{eff} \ge |p_{eff}|
    \end{equation}
    \item[$\bullet$] {\bf Strong energy condition (SEC):}
    \begin{equation}\label{Eq:sec}
        \rho_{eff} + 3p_{eff} \ge 0.
    \end{equation}
\end{itemize}


Using equations (\ref{Eq:M1reff}) - (\ref{Eq:M1peff}), (\ref{Eq:M2reff}) - (\ref{Eq:M2peff}) and (\ref{Eq:M3reff}) - (\ref{Eq:M3peff}) with equation (\ref{Eq:H}), we plot the inequalities (\ref{Eq:wec}) - (\ref{Eq:sec}) for the power-law model, Log-square root model and Exponential model, respectively. Moreover, we take the best-fit values given in Table \ref{table:2} for the Hubble parameter ($H_{0}$), jerk parameter ($j_{0}$) and deceleration parameter ($q_{0}$). 

\subsection{Power-law model}
\begin{figure}
\centering
\begin{subfigure}{.5\textwidth}
  \centering
  \includegraphics[width=6cm]{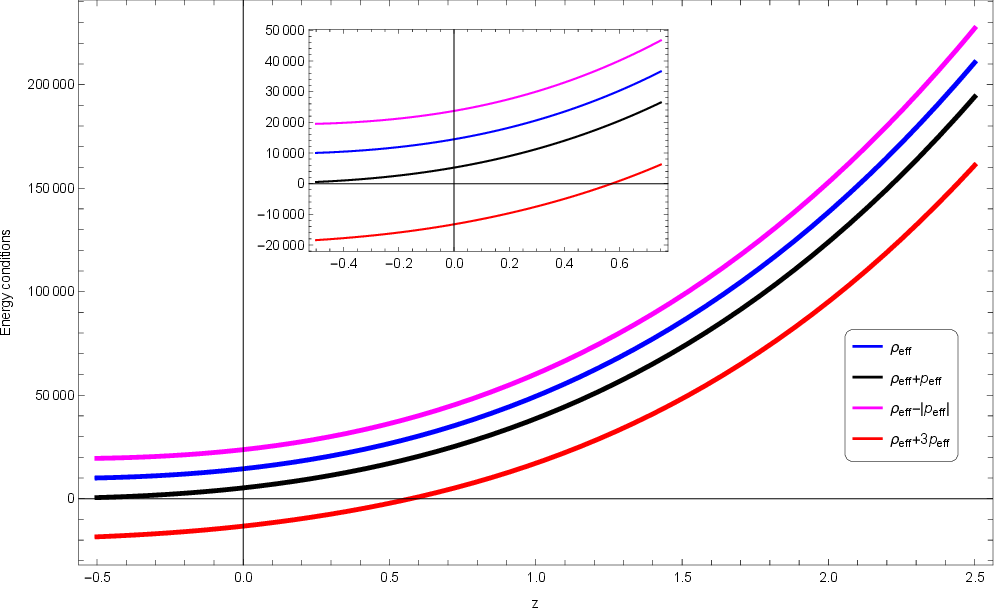}
  \caption{For $n= 0.33$}
  \label{m1n1}
\end{subfigure}%
\begin{subfigure}{.5\textwidth}
  \centering
  \includegraphics[width=6cm]{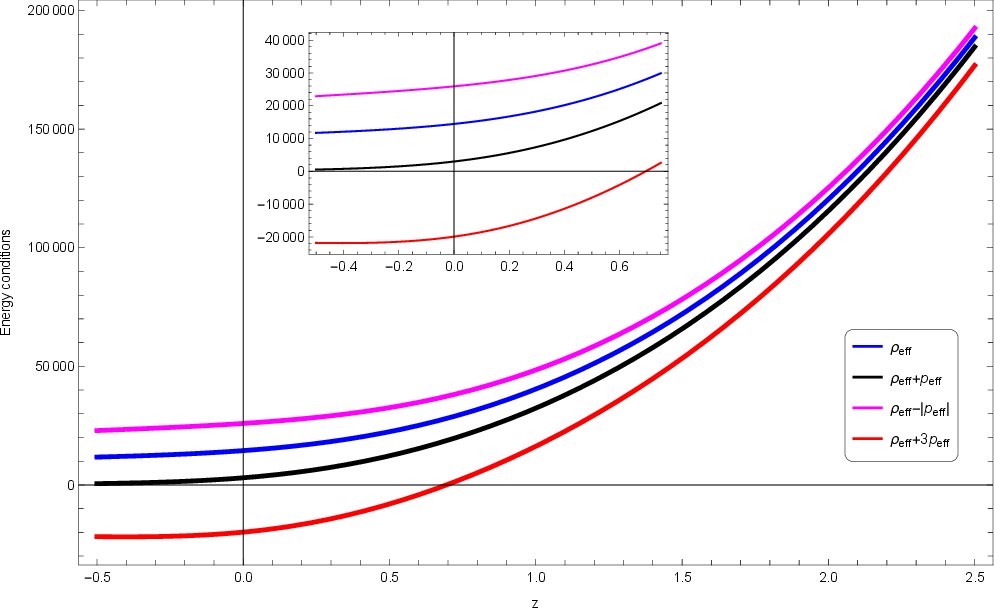}
  \caption{For $n = - 0.5$}
  \label{m1n2}
\end{subfigure}

\vspace{1cm}
\begin{subfigure}{.5\textwidth}
  \centering
  \includegraphics[width=6cm]{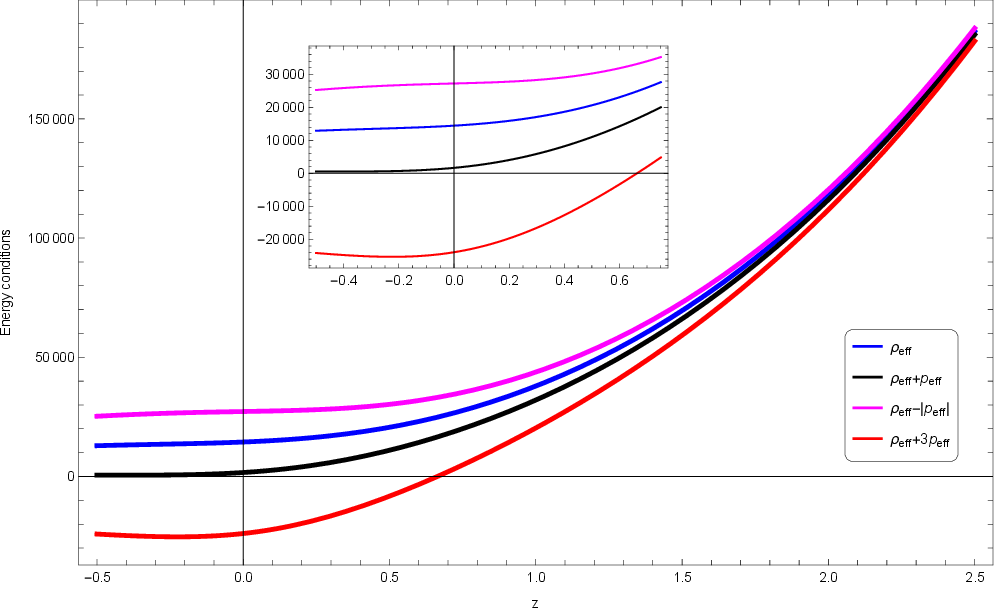}
  \caption{For $n = -1.0$}
  \label{m1n3}
\end{subfigure}
\caption{The schematic plot of energy conditions for the power-law model of $f(Q)$ gravity. The panel (a) for $n = 0.33$, (b) for $n = - 0.5$ and (c) for $n = - 1$ with the best fitting values of combined CC + pantheon data. The colors blue, black, red, and magenta shows the variations of $\rho_{eff}(z)$, $\rho_{eff}(z) + p_{eff}(z)$, $\rho_{eff}(z) - p_{eff}(z)$, and $\rho_{eff}(z) + 3p_{eff}(z)$ as a function of $z$.}
\label{ecs_m1} 
\end{figure}

In Fig. \ref{ecs_m1}, we show the schematic plot of $\rho_{eff}(z)$, $\rho_{eff}(z) + p_{eff}(z)$, $\rho_{eff}(z) - p_{eff}(z)$, and $\rho_{eff}(z) + 3p_{eff}(z)$ as a function of redshift parameter ($z$) for power-law model with different $n$ values $n = 0.33, - 0.5, -1.0$. In the figure, we observed that the above functions are decreasing as the universe expands, {\it i.e.}, decreasing of redshift $z$ .

\subsection{Log-square root model}
\begin{figure}[t]
\centering
\begin{subfigure}{.5\textwidth}
  \centering
  \includegraphics[width=6cm]{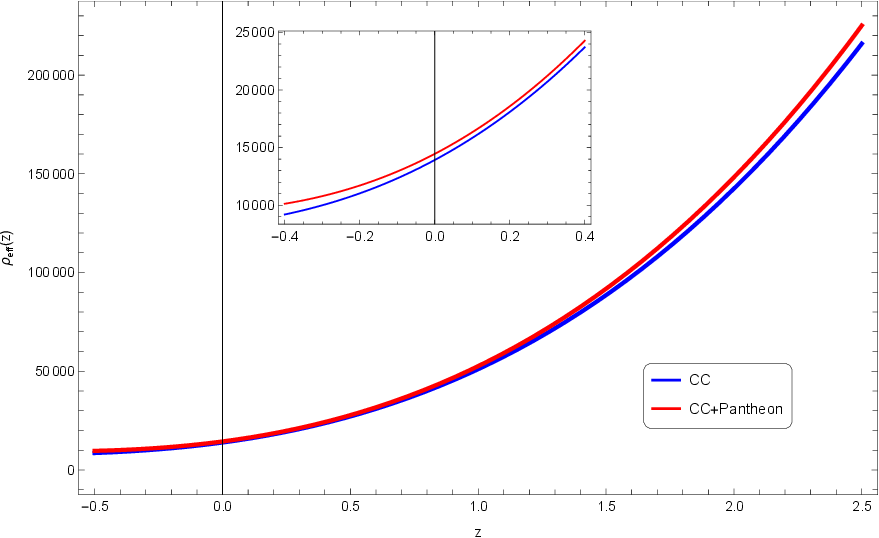}
  \caption{}
  \label{m2a}
\end{subfigure}%
\begin{subfigure}{.5\textwidth}
  \centering
  \includegraphics[width=6cm]{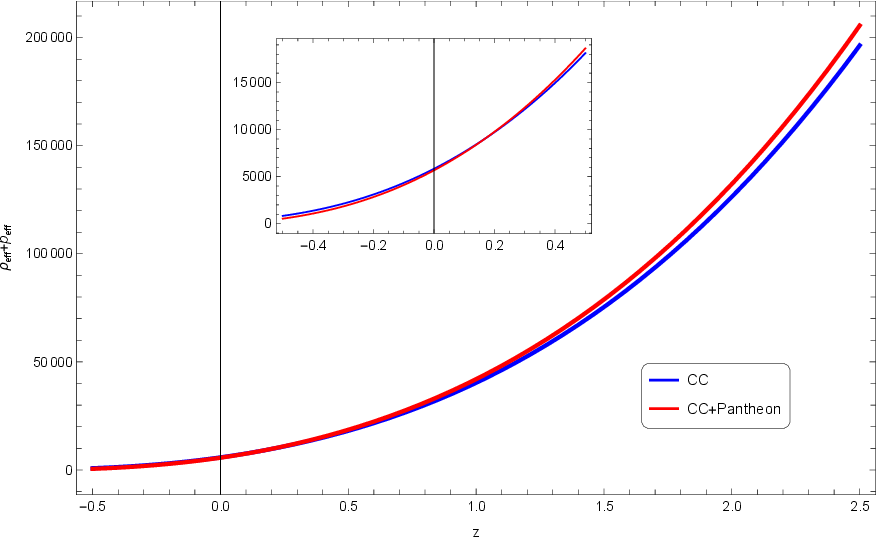}
  \caption{}
  \label{m2b}
\end{subfigure}

\vspace{1cm}
\begin{subfigure}{.5\textwidth}
  \centering
  \includegraphics[width=6cm]{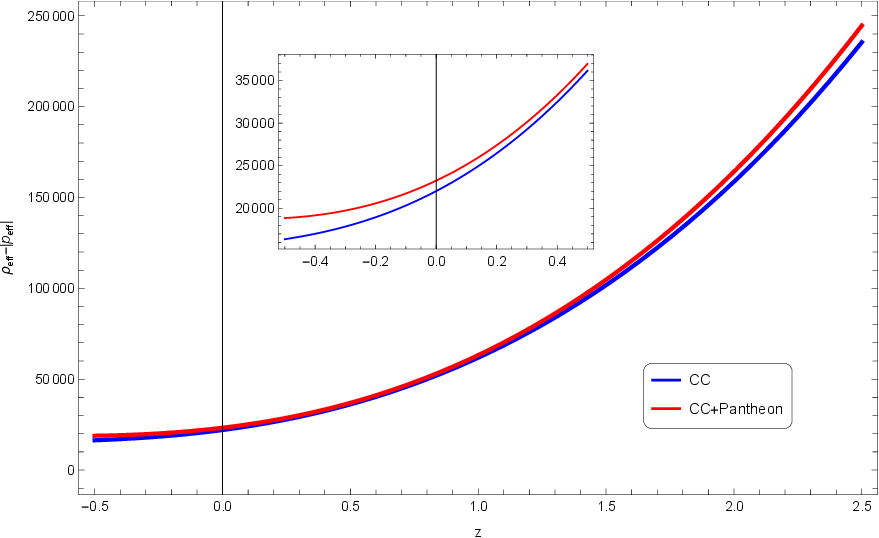}
  \caption{}
  \label{m2c}
\end{subfigure}%
\begin{subfigure}{.5\textwidth}
  \centering
  \includegraphics[width=6cm]{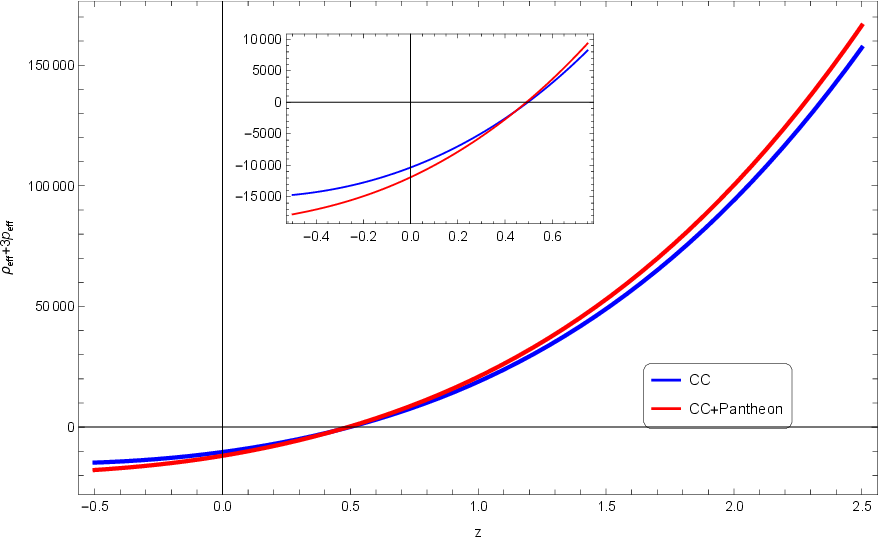}
  \caption{}
  \label{m2d}
\end{subfigure}
\caption{The schematic plot of energy conditions for the Log-square root model of $f(Q)$ gravity. The panels (a) $\rho_{eff}(z)$ (b) $\rho_{eff}(z) + p_{eff}(z)$, (c) $\rho_{eff}(z) - p_{eff}(z)$ and (d) $\rho_{eff}(z) + 3p_{eff}(z)$ as a function of $z$. The blue and red curves correspond to the best-fit values of CC and combined CC + Pantheon datasets.}
\label{ecs_m2} 
\end{figure}

In Fig. \ref{ecs_m2}, we show the schematic plot of (a) $\rho_{eff}(z)$ (b) $\rho_{eff}(z) + p_{eff}(z)$, (c) $\rho_{eff}(z) - p_{eff}(z)$ and (d) $\rho_{eff}(z) + 3p_{eff}(z)$ as a function of redshift parameter ($z$) for log-square root model for the best-fit values of CC and combined CC + Pantheon datasets. It is also noted that the $\rho_{eff}(z)$, $\rho_{eff}(z) + p_{eff}(z)$, $\rho_{eff}(z) - p_{eff}(z)$, and $\rho_{eff}(z) + 3p_{eff}(z)$ are the decreasing functions as the redshift $z$ decreases.  

\subsection{Exponential model}
\begin{figure}
\centering
\begin{subfigure}{.5\textwidth}
  \centering
  \includegraphics[width=6cm]{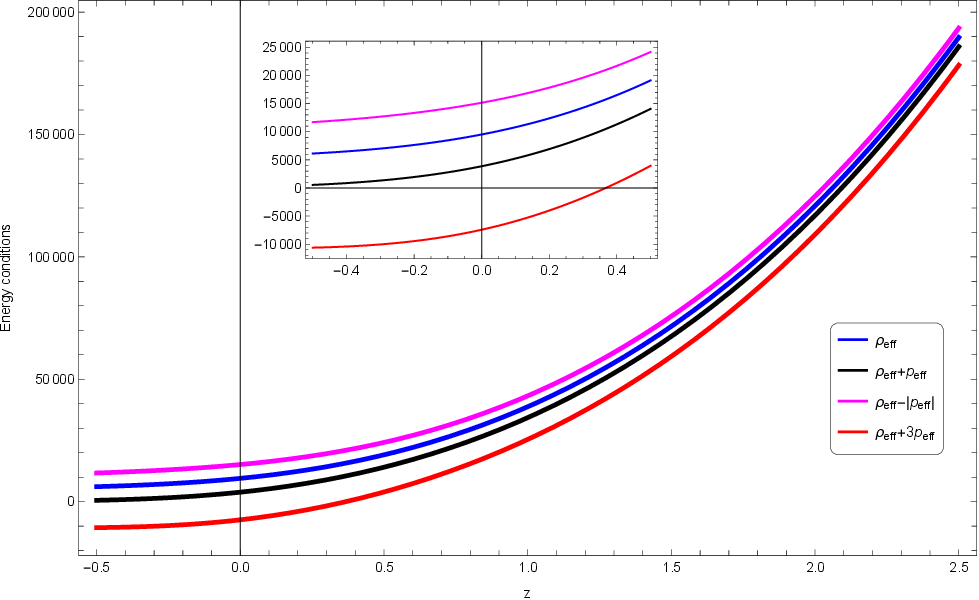}
  \caption{}
  \label{m3b1}
\end{subfigure}%
\begin{subfigure}{.5\textwidth}
  \centering
  \includegraphics[width=6cm]{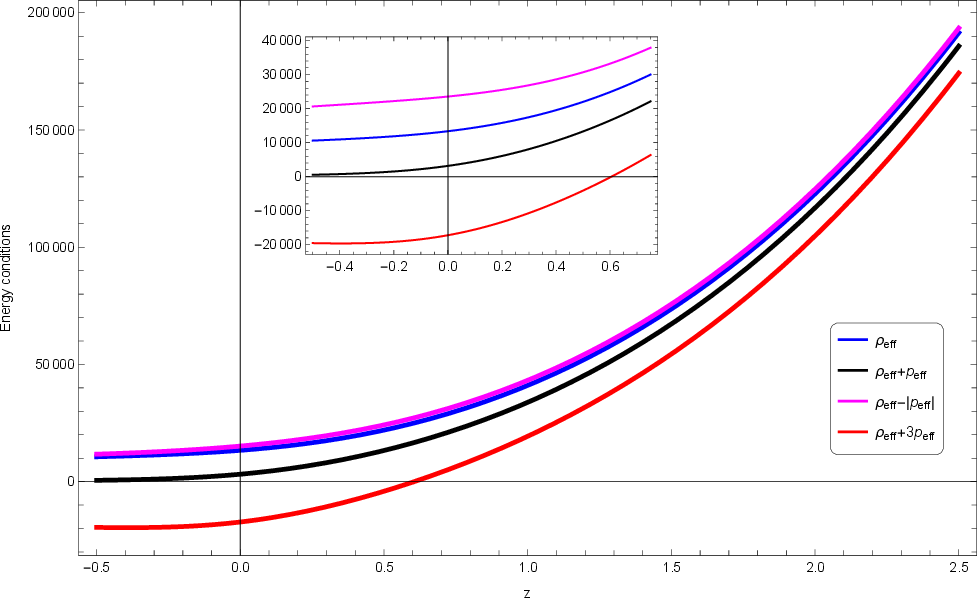}
  \caption{}
  \label{m3b2}
\end{subfigure}
\caption{The schematic plot of energy conditions for the exponential model of $f(Q)$ gravity. The panel (a) for $\beta = 0.25$ and (b) for $\beta = 0.37$ with the best fitting values of the combined CC + Pantheon data. The colours blue, black, red, and magenta represent the variation of $\rho_{eff}(z)$, $\rho_{eff}(z) + p_{eff}(z)$, $\rho_{eff}(z) - p_{eff}(z)$, and $\rho_{eff}(z) + 3p_{eff}(z)$.}
\label{ecs_m3} 
\end{figure}

The schematic plot of $\rho_{eff}(z)$, $\rho_{eff}(z) + p_{eff}(z)$, $\rho_{eff}(z) - p_{eff}(z)$, and $\rho_{eff}(z) + 3p_{eff}(z)$ as a function of redshift parameter ($z$) for the exponential model of $f(Q)$ gravity are shown in Fig. \ref{ecs_m3}. The left panel for $\beta = 0.25$ and the right panel for $\beta = 0.37$.

From these plots, we also observe that $\rho_{eff}(z)$, $\rho_{eff}(z) + p_{eff}(z)$ and $\rho_{eff}(z) - p_{eff}(z)$ remain positive throughout the universe's evolution, from early to present times for best-fit values of CC and CC+Pantheon datasets. However, $\rho_{eff}(z) + 3p_{eff}(z)$ becomes negative at a finite value of $z$, indicating a significant change in the universe's dynamics. This behaviour indicate an important implications for the energy conditions in cosmology:
\begin{itemize}
    \item Since $\rho_{eff}(z)$ and $\rho_{eff}(z) + p_{eff}(z)$ are positive, the WEC is satisfied.
    \item Given that $\rho_{eff}(z) + p_{eff}(z)$ is positive, the NEC is also satisfied.
    \item The positivity of $\rho_{eff}$ and $\rho_{eff}(z) - p_{eff}(z)$ ensures the DEC is satisfied.
    \item The fact that $\rho_{eff}(z) + 3p_{eff}(z)$ becomes negative at a certain redshift implies that the SEC is violated at later times.
\end{itemize}
The violation of the SEC in the late universe indicates that the universe is currently undergoing accelerated expansion. This accelerated expansion is consistent with observations of distant supernovae and the cosmic microwave background, which indicate the presence of DE or a cosmological constant driving this accelerated growth.

\section{Conclusions}
\label{sec:cls}
In the paper, we study the evolution of the late universe in the modified $f(Q)$ theory of gravity taking three different functional forms. As we do not know the correct form of $f(Q)$, we have employed three different forms of the modified gravity which are generally used to obtain cosmological models in the literature. We consider three different functional forms to construct cosmological models namely, (Model-I) power-law  $\left[f(Q) = Q + \alpha\left(\frac{Q}{Q_{0}}\right)^{n}\right]$, (Model-II) log-square-root  $\Big[f(Q) = Q + nQ_{0}\sqrt{\frac{Q}{\lambda Q_{0}}}\ln{\left(\frac{\lambda Q_{0}}{Q}\right)}\Big]$ and (Model-III)  exponential expansion $\Big[f(Q) = Qe^{\beta\frac{Q_{0}}{Q}}\Big]$ for the study. As the field equation is highly nonlinear  the dynamics of the late universe is investigated knowing a specific Hubble parameter which is determined from a constant jerk parameter. Late evolution is then studied expressing the Hubble parameter as a function of redshift parameter ( $z$ ). In section \ref{sec: obs}, we determine best-fit values of Hubble constant $H_{0}$, jerk parameter $j_{0}$ and deceleration parameter $q_{0}$ using the CC and combined analysis of CC + Pantheon data at the present epoch which are displayed in Table \ref{table:2}. The analysis of cosmological models with  CC + Pantheon is more accurate to observational predictions. The best-fit  Hubble parameter $H(z)$ and apparent magnitude $m(z)$ verses redshift $z$ are shown in Figs. \ref{fit:HD} and \ref{fit:SN}, respectively. The cosmological parameters namely, the deceleration and snap parameters are plotted in Figs. \ref{fig:dece_ccp} and \ref{fig:snap_ccp}. The deceleration parameter flips sign from positive ($q > 0$) to a negative value ($q < 0$) indicating a transition from decelerating phase in the past to accelerating phase at the present era. It is evident that such transition occurs later for the best-fit values of CC + Pantheon compared to the other data considered here. The snap parameter decreases as redshift decreases and at the present epoch it lies in the range (0, 1). It is predicted that the snap parameter may be negative ($s < 0$) in the future. The dynamical evolution of the density and EoS parameters as a function of the redshift parameter $z$ are plotted in Figs. \ref{M1rw} - \ref{M3rweff}. From the figures, it is evident that for power law model: The evolution of DE density and EoS parameter is studied for $n = -1$ using best-fit values of $H_{0}$, $j_{0}$ and $q_{0}$. The DE density $\rho_{DE}(z)$ increases as the universe expands, while the EoS parameter $\omega_{DE}(z)$ starts in the phantom region ($<-1$) at the present epoch and converges to $-1$ in the future, resembling a cosmological constant. The effective energy density $\rho_{eff}(z)$ remains positive, and the effective EoS parameter $\omega_{eff}(z)$ transitions from a decelerating phase ($\omega_{eff} > -\frac{1}{3}$) to an accelerating phase ($\omega_{eff} < -\frac{1}{3}$), with present values of $\omega_{eff} = - 0.89$ for CC and $\omega_{eff} = -0.94$ for CC+Pantheon, converging to $-1$ in the future. For different $n$ values, $\omega_{DE}$ converges to $-1$ eventually, with DE behaving as quintessence for $n > 0$ or phantom $n < 0$ at late times before transitioning to a cosmological constant-like state. For the log-square root model: DE density decreases with redshift but remains positive throughout cosmic evolution. The EoS parameter $\omega_{DE}(z)$ demonstrates quintessence behaviour ($- 1 < \omega_{DE} < - \frac{1}{3}$) from the early universe to the present. Using CC data, $\omega_{DE} \geq -1$ indicates DE initially behaves dynamically quintessence but transitions to act as a cosmological constant at late times. The CC + Pantheon data, however, suggests $\omega_{DE} < -1$, implying a universe dominated by phantom energy at late times. The effective energy density $\rho_{eff}(z)$ decreases with the universe's expansion while remaining positive. The effective EoS parameter $\omega_{eff}(z)$ shows the universe transitions from a decelerating phase ($\omega_{eff} > - \frac{1}{3}$) to an accelerating phase ($\omega_{eff} < - \frac{1}{3}$). In the future, $\omega_{eff}$ converges to $-1$, indicating a cosmological constant-like behaviour. For Exponential model: Using  best-fit values with $\beta =0.37$, DE density $\rho_{DE}(z)$ is plotted, it is evident that  DE density increases  as the universe expands. The initial EoS parameter $\omega_{DE}(z) < -1$  is found to decrease  to a minimum, thereafter it  increases. In the case of CC data, it  predicts a persistent phantom regime ($\omega_{DE} < -1$), while CC+Pantheon data indicates a transition from phantom to quintessence ($\omega_{DE}>-1$) in the future. The effective energy density $\rho_{eff}(z)$ remains positive, and the effective EoS paramter $\omega_{eff}(z)$ is found to decrease as the universe evolves.

The energy conditions are analyzed by plotting in Figs. \ref{ecs_m1} - \ref{ecs_m3} for the power-law model, log-square root model, and exponential model of the modified $F(Q)$ gravity respectively. The universe is found to transit from a decelerating phase to an accelerating phase in all the three models described here are ascertained but the time of flip of sign of SEC is found to depend on the model parameters implies that during accelerating phase the SEC is violated. This transition, which is a key feature for all the functional forms of $F(Q)$ gravity emphasize the critical role of DE for describing the cosmic acceleration. Even though the contributing fluids in each model differs, the role of DE plays a significant contribution for understanding the dynamics of the late universe predicting that it is a key component in overcoming the deceleration to an accelerating universe. Finally, we note the following for DE fluid behaviour: (i)  the power-law model admits both the phantom and quintessence fluid when  (a) $n<0$ and (b) $n>0$ respectively, (ii) quientessence only in log-square root model and (iii) phantom fluid only for exponential model.

\backmatter
\bmhead{Acknowledgements}

BCR and BCP express their sincere gratitude to the IUCAA Center for Astronomy Research and Development (ICARD), Department of Physics, North Bengal University, for providing invaluable research facilities. BCR also gratefully acknowledges the support of the Ministry of Social Justice and Empowerment, the Government of India, and the University Grants Commission (UGC), India, for the fellowship awarded during this research. BCP would like to thank SERB DST Govt. of India for a project grant(F. No. CRG/2021/000183).









\bibliography{sn-bibliography}


\begin{thebibliography}{111}
\ifx \bisbn   \undefined \def \bisbn  #1{ISBN #1}\fi
\ifx \binits  \undefined \def \binits#1{#1}\fi
\ifx \bauthor  \undefined \def \bauthor#1{#1}\fi
\ifx \batitle  \undefined \def \batitle#1{#1}\fi
\ifx \bjtitle  \undefined \def \bjtitle#1{#1}\fi
\ifx \bvolume  \undefined \def \bvolume#1{\textbf{#1}}\fi
\ifx \byear  \undefined \def \byear#1{#1}\fi
\ifx \bissue  \undefined \def \bissue#1{#1}\fi
\ifx \bfpage  \undefined \def \bfpage#1{#1}\fi
\ifx \blpage  \undefined \def \blpage #1{#1}\fi
\ifx \burl  \undefined \def \burl#1{\textsf{#1}}\fi
\ifx \doiurl  \undefined \def \doiurl#1{\url{https://doi.org/#1}}\fi
\ifx \betal  \undefined \def \betal{\textit{et al.}}\fi
\ifx \binstitute  \undefined \def \binstitute#1{#1}\fi
\ifx \binstitutionaled  \undefined \def \binstitutionaled#1{#1}\fi
\ifx \bctitle  \undefined \def \bctitle#1{#1}\fi
\ifx \beditor  \undefined \def \beditor#1{#1}\fi
\ifx \bpublisher  \undefined \def \bpublisher#1{#1}\fi
\ifx \bbtitle  \undefined \def \bbtitle#1{#1}\fi
\ifx \bedition  \undefined \def \bedition#1{#1}\fi
\ifx \bseriesno  \undefined \def \bseriesno#1{#1}\fi
\ifx \blocation  \undefined \def \blocation#1{#1}\fi
\ifx \bsertitle  \undefined \def \bsertitle#1{#1}\fi
\ifx \bsnm \undefined \def \bsnm#1{#1}\fi
\ifx \bsuffix \undefined \def \bsuffix#1{#1}\fi
\ifx \bparticle \undefined \def \bparticle#1{#1}\fi
\ifx \barticle \undefined \def \barticle#1{#1}\fi
\bibcommenthead
\ifx \bconfdate \undefined \def \bconfdate #1{#1}\fi
\ifx \botherref \undefined \def \botherref #1{#1}\fi
\ifx \url \undefined \def \url#1{\textsf{#1}}\fi
\ifx \bchapter \undefined \def \bchapter#1{#1}\fi
\ifx \bbook \undefined \def \bbook#1{#1}\fi
\ifx \bcomment \undefined \def \bcomment#1{#1}\fi
\ifx \oauthor \undefined \def \oauthor#1{#1}\fi
\ifx \citeauthoryear \undefined \def \citeauthoryear#1{#1}\fi
\ifx \endbibitem  \undefined \def \endbibitem {}\fi
\ifx \bconflocation  \undefined \def \bconflocation#1{#1}\fi
\ifx \arxivurl  \undefined \def \arxivurl#1{\textsf{#1}}\fi
\csname PreBibitemsHook\endcsname

\bibitem[\protect\citeauthoryear{{Perlmutter} et~al.}{1999}]{1999ApJ...517..565P}
\begin{barticle}
\bauthor{\bsnm{{Perlmutter}}, \binits{S.}}, \betal:
\batitle{{Measurements of {\ensuremath{\Omega}} and {\ensuremath{\Lambda}} from 42 High-Redshift Supernovae}}.
\bjtitle{Astrophy. J.}
\bvolume{517}(\bissue{2}),
\bfpage{565}--\blpage{586}
(\byear{1999})
\doiurl{10.1086/307221}
{\href{https://arxiv.org/abs/astro-ph/9812133}{{arXiv:astro-ph/9812133}}}
\end{barticle}
\endbibitem

\bibitem[\protect\citeauthoryear{Riess et~al.}{1998}]{riess_observational_1998}
\begin{barticle}
\bauthor{\bsnm{Riess}, \binits{A.G.}}, \betal:
\batitle{{Observational Evidence from Supernovae for an Accelerating Universe and a Cosmological Constant}}.
\bjtitle{{Astron. J.}}
\bvolume{116}(\bissue{3}),
\bfpage{1009}
(\byear{1998})
\doiurl{10.1086/300499}
\end{barticle}
\endbibitem

\bibitem[\protect\citeauthoryear{Riess et~al.}{2004}]{riess2004type}
\begin{barticle}
\bauthor{\bsnm{Riess}, \binits{A.G.}}, \betal:
\batitle{{Type Ia supernova discoveries at {\ensuremath{z > 1}} from the Hubble Space Telescope: Evidence for past deceleration and constraints on dark energy evolution}}.
\bjtitle{Astrophy. J.}
\bvolume{607}(\bissue{2}),
\bfpage{665}
(\byear{2004})
\doiurl{10.1086/383612}
\end{barticle}
\endbibitem

\bibitem[\protect\citeauthoryear{Spergel et~al.}{2003}]{Spergel_2003}
\begin{barticle}
\bauthor{\bsnm{Spergel}, \binits{D.N.}}, \betal:
\batitle{{First Year Wilkinson Microwave Anisotropy Probe (WMAP) Observations: Determination of Cosmological Parameters}}.
\bjtitle{Astrophys. J. Suppl. Series}
\bvolume{148}(\bissue{1}),
\bfpage{175}--\blpage{194}
(\byear{2003})
\doiurl{10.1086/377226}
\end{barticle}
\endbibitem

\bibitem[\protect\citeauthoryear{Koivisto and Mota}{2006}]{Koivisto_2006}
\begin{botherref}
\oauthor{\bsnm{Koivisto}, \binits{T.}},
\oauthor{\bsnm{Mota}, \binits{D.F.}}:
{Dark energy anisotropic stress and large scale structure formation}.
Phys. Rev. D
\textbf{73}(8)
(2006)
\doiurl{10.1103/physrevd.73.083502}
\end{botherref}
\endbibitem

\bibitem[\protect\citeauthoryear{Daniel et~al.}{2008}]{Daniel_2008}
\begin{botherref}
\oauthor{\bsnm{Daniel}, \binits{S.F.}},
\oauthor{\bsnm{Caldwell}, \binits{R.R.}},
\oauthor{\bsnm{Cooray}, \binits{A.}},
\oauthor{\bsnm{Melchiorri}, \binits{A.}}:
{Large scale structure as a probe of gravitational slip}.
Phys. Rev. D
\textbf{77}(10)
(2008)
\doiurl{10.1103/physrevd.77.103513}
\end{botherref}
\endbibitem

\bibitem[\protect\citeauthoryear{Minami and Komatsu}{2020}]{Minami_2020}
\begin{botherref}
\oauthor{\bsnm{Minami}, \binits{Y.}},
\oauthor{\bsnm{Komatsu}, \binits{E.}}:
{New Extraction of the Cosmic Birefringence from the Planck 2018 Polarization Data}.
Phys. Rev. Lett.
\textbf{125}(22)
(2020)
\doiurl{10.1103/physrevlett.125.221301}
\end{botherref}
\endbibitem

\bibitem[\protect\citeauthoryear{Sahni and Starobinsky}{2000}]{SAHNI_2000}
\begin{barticle}
\bauthor{\bsnm{Sahni}, \binits{V.}},
\bauthor{\bsnm{Starobinsky}, \binits{A.}}:
\batitle{{THE CASE FOR A POSITIVE COSMOLOGICAL $\Lambda$-TERM}}.
\bjtitle{Int. J. Mod. Phys. D}
\bvolume{09}(\bissue{04}),
\bfpage{373}--\blpage{443}
(\byear{2000})
\doiurl{10.1142/s0218271800000542}
\end{barticle}
\endbibitem

\bibitem[\protect\citeauthoryear{Carroll}{2001}]{Carroll_2001}
\begin{botherref}
\oauthor{\bsnm{Carroll}, \binits{S.M.}}:
The cosmological constant.
Living Rev. Rel.
\textbf{4}(1)
(2001)
\doiurl{10.12942/lrr-2001-1}
\end{botherref}
\endbibitem

\bibitem[\protect\citeauthoryear{Padmanabhan}{2003}]{Padmanabhan_2003}
\begin{barticle}
\bauthor{\bsnm{Padmanabhan}, \binits{T.}}:
\batitle{Cosmological constant{\textemdash}the weight of the vacuum}.
\bjtitle{Phys. Rept.}
\bvolume{380}(\bissue{5-6}),
\bfpage{235}--\blpage{320}
(\byear{2003})
\doiurl{10.1016/s0370-1573(03)00120-0}
\end{barticle}
\endbibitem

\bibitem[\protect\citeauthoryear{Peebles and Ratra}{2003}]{Peebles_2003}
\begin{barticle}
\bauthor{\bsnm{Peebles}, \binits{P.J.E.}},
\bauthor{\bsnm{Ratra}, \binits{B.}}:
\batitle{The cosmological constant and dark energy}.
\bjtitle{Rev. Mod. Phys.}
\bvolume{75}(\bissue{2}),
\bfpage{559}--\blpage{606}
(\byear{2003})
\doiurl{10.1103/revmodphys.75.559}
\end{barticle}
\endbibitem

\bibitem[\protect\citeauthoryear{Chiba}{1999}]{chiba_quintessence_1999}
\begin{barticle}
\bauthor{\bsnm{Chiba}, \binits{T.}}:
\batitle{Quintessence, the gravitational constant, and gravity}.
\bjtitle{Phys. Rev. D}
\bvolume{60}(\bissue{8}),
\bfpage{083508}
(\byear{1999})
\doiurl{10.1103/PhysRevD.60.083508}
\end{barticle}
\endbibitem

\bibitem[\protect\citeauthoryear{Amendola}{2000}]{amendola_coupled_2000}
\begin{barticle}
\bauthor{\bsnm{Amendola}, \binits{L.}}:
\batitle{Coupled quintessence}.
\bjtitle{Phys. Rev. D}
\bvolume{62}(\bissue{4}),
\bfpage{043511}
(\byear{2000})
\doiurl{10.1103/PhysRevD.62.043511}
\end{barticle}
\endbibitem

\bibitem[\protect\citeauthoryear{Martin}{2008}]{martin_quintessence_2008}
\begin{barticle}
\bauthor{\bsnm{Martin}, \binits{J.}}:
\batitle{Quintessence: a mini-review}.
\bjtitle{Mod. Phys. Lett. A}
\bvolume{23}(\bissue{17}),
\bfpage{1252}--\blpage{1265}
(\byear{2008})
\doiurl{10.1142/S0217732308027631} .
\bcomment{Publisher: World Scientific Publishing Co.}
Accessed 2023-05-24
\end{barticle}
\endbibitem

\bibitem[\protect\citeauthoryear{Kamenshchik et~al.}{2001}]{Kamenshchik_2001}
\begin{barticle}
\bauthor{\bsnm{Kamenshchik}, \binits{A.}},
\bauthor{\bsnm{Moschella}, \binits{U.}},
\bauthor{\bsnm{Pasquier}, \binits{V.}}:
\batitle{An alternative to quintessence}.
\bjtitle{Phys. Lett. B}
\bvolume{511}(\bissue{2-4}),
\bfpage{265}--\blpage{268}
(\byear{2001})
\doiurl{10.1016/s0370-2693(01)00571-8}
\end{barticle}
\endbibitem

\bibitem[\protect\citeauthoryear{Bento et~al.}{2002}]{Bento_2002}
\begin{botherref}
\oauthor{\bsnm{Bento}, \binits{M.C.}},
\oauthor{\bsnm{Bertolami}, \binits{O.}},
\oauthor{\bsnm{Sen}, \binits{A.A.}}:
Generalized chaplygin gas, accelerated expansion, and dark-energy matter unification.
Phys. Rev. D
\textbf{66}(4)
(2002)
\doiurl{10.1103/physrevd.66.043507}
\end{botherref}
\endbibitem

\bibitem[\protect\citeauthoryear{Benaoum}{2002}]{Benaoum_2002}
\begin{barticle}
\bauthor{\bsnm{Benaoum}, \binits{H.}}:
\batitle{Accelerated universe from modified chaplygin gas and tachyonic fluid}.
\bjtitle{Universe}
\bvolume{8}(\bissue{7}),
\bfpage{340}
(\byear{2002})
\doiurl{10.3390/universe8070340}
\end{barticle}
\endbibitem

\bibitem[\protect\citeauthoryear{Sotiriou and Faraoni}{2010}]{sotiriou_fr_2010}
\begin{barticle}
\bauthor{\bsnm{Sotiriou}, \binits{T.P.}},
\bauthor{\bsnm{Faraoni}, \binits{V.}}:
\batitle{{$f(R)$ theories of gravity}}.
\bjtitle{Rev. Mod. Phys.}
\bvolume{82}(\bissue{1}),
\bfpage{451}--\blpage{497}
(\byear{2010})
\doiurl{10.1103/RevModPhys.82.451}
\end{barticle}
\endbibitem

\bibitem[\protect\citeauthoryear{Felice and Tsujikawa}{2010}]{De_Felice_2010}
\begin{botherref}
\oauthor{\bsnm{Felice}, \binits{A.D.}},
\oauthor{\bsnm{Tsujikawa}, \binits{S.}}:
{$f(R)$ Theories}.
Living Rev. Rel.
\textbf{13}(1)
(2010)
\doiurl{10.12942/lrr-2010-3}
\end{botherref}
\endbibitem

\bibitem[\protect\citeauthoryear{Harko et~al.}{2011}]{Harko_2011}
\begin{botherref}
\oauthor{\bsnm{Harko}, \binits{T.}},
\oauthor{\bsnm{Lobo}, \binits{F.S.N.}},
\oauthor{\bsnm{Nojiri}, \binits{S.}},
\oauthor{\bsnm{Odintsov}, \binits{S.D.}}:
{$f(R,T)$ gravity}.
Phys. Rev. D
\textbf{84}(2)
(2011)
\doiurl{10.1103/physrevd.84.024020}
\end{botherref}
\endbibitem

\bibitem[\protect\citeauthoryear{Nojiri and Odintsov}{2005}]{NOJIRI20051}
\begin{barticle}
\bauthor{\bsnm{Nojiri}, \binits{S.}},
\bauthor{\bsnm{Odintsov}, \binits{S.D.}}:
\batitle{{Modified Gauss-Bonnet theory as gravitational alternative for dark energy}}.
\bjtitle{Phys. Lett. B}
\bvolume{631}(\bissue{1}),
\bfpage{1}--\blpage{6}
(\byear{2005})
\doiurl{10.1016/j.physletb.2005.10.010}
\end{barticle}
\endbibitem

\bibitem[\protect\citeauthoryear{Cognola et~al.}{2006}]{PhysRevD.73.084007}
\begin{barticle}
\bauthor{\bsnm{Cognola}, \binits{G.}},
\bauthor{\bsnm{Elizalde}, \binits{E.}},
\bauthor{\bsnm{Nojiri}, \binits{S.}},
\bauthor{\bsnm{Odintsov}, \binits{S.D.}},
\bauthor{\bsnm{Zerbini}, \binits{S.}}:
\batitle{{Dark energy in modified Gauss-Bonnet gravity: Late-time acceleration and the hierarchy problem}}.
\bjtitle{Phys. Rev. D}
\bvolume{73},
\bfpage{084007}
(\byear{2006})
\doiurl{10.1103/PhysRevD.73.084007}
\end{barticle}
\endbibitem

\bibitem[\protect\citeauthoryear{Li et~al.}{2007}]{PhysRevD.76.044027}
\begin{barticle}
\bauthor{\bsnm{Li}, \binits{B.}},
\bauthor{\bsnm{Barrow}, \binits{J.D.}},
\bauthor{\bsnm{Mota}, \binits{D.F.}}:
\batitle{{Cosmology of modified Gauss-Bonnet gravity}}.
\bjtitle{Phys. Rev. D}
\bvolume{76},
\bfpage{044027}
(\byear{2007})
\doiurl{10.1103/PhysRevD.76.044027}
\end{barticle}
\endbibitem

\bibitem[\protect\citeauthoryear{Maartens and Koyama}{2010}]{Maartens_2010}
\begin{botherref}
\oauthor{\bsnm{Maartens}, \binits{R.}},
\oauthor{\bsnm{Koyama}, \binits{K.}}:
Brane-world gravity.
Living Rev. Rel.
\textbf{13}(1)
(2010)
\doiurl{10.12942/lrr-2010-5}
\end{botherref}
\endbibitem

\bibitem[\protect\citeauthoryear{Brax et~al.}{2004}]{Brax_2004}
\begin{barticle}
\bauthor{\bsnm{Brax}, \binits{P.}},
\bauthor{\bsnm{Bruck}, \binits{C.}},
\bauthor{\bsnm{Davis}, \binits{A.-C.}}:
\batitle{Brane world cosmology}.
\bjtitle{Rept. Prog. Phys.}
\bvolume{67}(\bissue{12}),
\bfpage{2183}--\blpage{2231}
(\byear{2004})
\doiurl{10.1088/0034-4885/67/12/r02}
\end{barticle}
\endbibitem

\bibitem[\protect\citeauthoryear{Wang}{2017}]{wang_horava_2017}
\begin{barticle}
\bauthor{\bsnm{Wang}, \binits{A.}}:
\batitle{Hořava gravity at a lifshitz point: A progress report}.
\bjtitle{Int. J. Mod. Phys. D}
\bvolume{26}(\bissue{7}),
\bfpage{1730014}
(\byear{2017})
\doiurl{10.1142/S0218271817300142}
\end{barticle}
\endbibitem

\bibitem[\protect\citeauthoryear{Starobinsky}{1980}]{starobinsky198030}
\begin{botherref}
\oauthor{\bsnm{Starobinsky}, \binits{A.}}:
A new type of isotropic cosmological models without singularity.
Phys. Lett. B
\textbf{99}
(1980)
\doiurl{10.1016/0370-2693(80)90670-X}
\end{botherref}
\endbibitem

\bibitem[\protect\citeauthoryear{Capozziello et~al.}{2008}]{PhysRevD.78.063504}
\begin{barticle}
\bauthor{\bsnm{Capozziello}, \binits{S.}},
\bauthor{\bsnm{Cardone}, \binits{V.F.}},
\bauthor{\bsnm{Salzano}, \binits{V.}}:
\batitle{{Cosmography of $f(R)$ gravity}}.
\bjtitle{Phys. Rev. D}
\bvolume{78},
\bfpage{063504}
(\byear{2008})
\doiurl{10.1103/PhysRevD.78.063504}
\end{barticle}
\endbibitem

\bibitem[\protect\citeauthoryear{Capozziello and Laurentis}{2012}]{Capozziello_2012_AdP}
\begin{barticle}
\bauthor{\bsnm{Capozziello}, \binits{S.}},
\bauthor{\bsnm{Laurentis}, \binits{M.D.}}:
\batitle{{The dark matter problem from $f(R)$ gravity viewpoint}}.
\bjtitle{Ann. der Phys.}
\bvolume{524},
\bfpage{545}--\blpage{578}
(\byear{2012})
\doiurl{10.1002/andp.201200109}
\end{barticle}
\endbibitem

\bibitem[\protect\citeauthoryear{Myrzakulov}{2012}]{Myrzakulov_2012}
\begin{botherref}
\oauthor{\bsnm{Myrzakulov}, \binits{R.}}:
{FRW cosmology in $F(R,T)$ gravity}.
Euro. Phys. J. C
\textbf{72}(11)
(2012)
\doiurl{10.1140/epjc/s10052-012-2203-y}
\end{botherref}
\endbibitem

\bibitem[\protect\citeauthoryear{Rudra and Giri}{2021}]{RUDRA2021115428}
\begin{barticle}
\bauthor{\bsnm{Rudra}, \binits{P.}},
\bauthor{\bsnm{Giri}, \binits{K.}}:
\batitle{{Observational constraint in $f(R,T)$ gravity from the cosmic chronometers and some standard distance measurement parameters}}.
\bjtitle{Nucl. Phys. B}
\bvolume{967},
\bfpage{115428}
(\byear{2021})
\doiurl{10.1016/j.nuclphysb.2021.115428}
\end{barticle}
\endbibitem

\bibitem[\protect\citeauthoryear{Paul et~al.}{2022}]{Paul_2022}
\begin{barticle}
\bauthor{\bsnm{Paul}, \binits{B.C.}},
\bauthor{\bsnm{Chanda}, \binits{A.}},
\bauthor{\bsnm{Beesham}, \binits{A.}},
\bauthor{\bsnm{Maharaj}, \binits{S.D.}}:
\batitle{{Late time cosmology in $f(R,\mathcal{G})$-gravity with interacting fluids}}.
\bjtitle{Class. Quantum Grav.}
\bvolume{39}(\bissue{6}),
\bfpage{065006}
(\byear{2022})
\doiurl{10.1088/1361-6382/ac4b97}
\end{barticle}
\endbibitem

\bibitem[\protect\citeauthoryear{Nojiri et~al.}{2006}]{Nojiri_2006}
\begin{barticle}
\bauthor{\bsnm{Nojiri}, \binits{S.}},
\bauthor{\bsnm{Odintsov}, \binits{S.D.}},
\bauthor{\bsnm{Gorbunova}, \binits{O.G.}}:
\batitle{{Dark energy problem: from phantom theory to modified Gauss{\textendash}Bonnet gravity}}.
\bjtitle{J. Phys. A: Mathematical and General}
\bvolume{39}(\bissue{21}),
\bfpage{6627}--\blpage{6633}
(\byear{2006})
\doiurl{10.1088/0305-4470/39/21/s62}
\end{barticle}
\endbibitem

\bibitem[\protect\citeauthoryear{Valentino et~al.}{2023}]{Di_Valentino_2023}
\begin{barticle}
\bauthor{\bsnm{Valentino}, \binits{E.D.}},
\bauthor{\bsnm{Nilsson}, \binits{N.A.}},
\bauthor{\bsnm{Park}, \binits{M.-I.}}:
\batitle{A new test of dynamical dark energy models and cosmic tensions in ho{\v{r} }ava gravity}.
\bjtitle{MNRAS}
\bvolume{519}(\bissue{4}),
\bfpage{5043}--\blpage{5058}
(\byear{2023})
\doiurl{10.1093/mnras/stac3824}
\end{barticle}
\endbibitem

\bibitem[\protect\citeauthoryear{Zhang et~al.}{2020}]{Zhang_2020}
\begin{botherref}
\oauthor{\bsnm{Zhang}, \binits{T.}},
\oauthor{\bsnm{Shu}, \binits{F.-W.}},
\oauthor{\bsnm{Tang}, \binits{Q.-W.}},
\oauthor{\bsnm{Du}, \binits{D.-H.}}:
Constraints on ho{\v{r}}ava{\textendash}lifshitz gravity from {GRB} 170817a.
Euro. Phys. J. C
\textbf{80}(11)
(2020)
\doiurl{10.1140/epjc/s10052-020-08626-z}
\end{botherref}
\endbibitem

\bibitem[\protect\citeauthoryear{Pellegrini and Plebanski}{1963}]{pellegrini1963tetrad}
\begin{botherref}
\oauthor{\bsnm{Pellegrini}, \binits{C.}},
\oauthor{\bsnm{Plebanski}, \binits{J.}}:
{Tetrad fields and gravitational fields}.
Kgl. Danske Videnskab. Selskab, Mat. Fys. Skrifter
\textbf{2}(4)
(1963)
\end{botherref}
\endbibitem

\bibitem[\protect\citeauthoryear{Hayashi and Shirafuji}{1979}]{PhysRevD.19.3524}
\begin{barticle}
\bauthor{\bsnm{Hayashi}, \binits{K.}},
\bauthor{\bsnm{Shirafuji}, \binits{T.}}:
\batitle{{New general relativity}}.
\bjtitle{Phys. Rev. D}
\bvolume{19},
\bfpage{3524}--\blpage{3553}
(\byear{1979})
\doiurl{10.1103/PhysRevD.19.3524}
\end{barticle}
\endbibitem

\bibitem[\protect\citeauthoryear{Linder}{2010}]{Linder_2010}
\begin{botherref}
\oauthor{\bsnm{Linder}, \binits{E.V.}}:
{Einstein's other gravity and the acceleration of the Universe}.
Phys. Rev. D
\textbf{81}(12)
(2010)
\doiurl{10.1103/physrevd.81.127301}
\end{botherref}
\endbibitem

\bibitem[\protect\citeauthoryear{Maluf}{2013}]{Maluf_2013}
\begin{barticle}
\bauthor{\bsnm{Maluf}, \binits{J.W.}}:
\batitle{{The teleparallel equivalent of general relativity}}.
\bjtitle{Ann. der Phys.}
\bvolume{525}(\bissue{5}),
\bfpage{339}--\blpage{357}
(\byear{2013})
\doiurl{10.1002/andp.201200272}
\end{barticle}
\endbibitem

\bibitem[\protect\citeauthoryear{Aldrovandi and Pereira}{2013}]{Aldrovandi:2013wha}
\begin{bbook}
\bauthor{\bsnm{Aldrovandi}, \binits{R.}},
\bauthor{\bsnm{Pereira}, \binits{J.G.}}:
\bbtitle{{Teleparallel Gravity: An Introduction}}
vol. \bseriesno{173}.
\bpublisher{Springer},
\blocation{New {Y}ork}
(\byear{2013}).
\doiurl{10.1007/978-94-007-5143-9}
\end{bbook}
\endbibitem

\bibitem[\protect\citeauthoryear{Capozziello et~al.}{2011}]{Capozziello_2011}
\begin{botherref}
\oauthor{\bsnm{Capozziello}, \binits{S.}},
\oauthor{\bsnm{Cardone}, \binits{V.F.}},
\oauthor{\bsnm{Farajollahi}, \binits{H.}},
\oauthor{\bsnm{Ravanpak}, \binits{A.}}:
{Cosmography in {\ensuremath{f(T)}} gravity}.
Phys. Rev. D
\textbf{84}(4)
(2011)
\doiurl{10.1103/physrevd.84.043527}
\end{botherref}
\endbibitem

\bibitem[\protect\citeauthoryear{Ferraro and Fiorini}{2007}]{PhysRevD.75.084031}
\begin{barticle}
\bauthor{\bsnm{Ferraro}, \binits{R.}},
\bauthor{\bsnm{Fiorini}, \binits{F.}}:
\batitle{Modified teleparallel gravity: Inflation without an inflaton}.
\bjtitle{Phys. Rev. D}
\bvolume{75},
\bfpage{084031}
(\byear{2007})
\doiurl{10.1103/PhysRevD.75.084031}
\end{barticle}
\endbibitem

\bibitem[\protect\citeauthoryear{Ferraro and Fiorini}{2008}]{ferraro_born-infeld_2008}
\begin{barticle}
\bauthor{\bsnm{Ferraro}, \binits{R.}},
\bauthor{\bsnm{Fiorini}, \binits{F.}}:
\batitle{{Born-Infeld gravity in Weitzenb{\"{o}}ck spacetime}}.
\bjtitle{Phys. Rev. D}
\bvolume{78}(\bissue{12}),
\bfpage{124019}
(\byear{2008})
\doiurl{10.1103/PhysRevD.78.124019} .
Accessed 2023-05-24
\end{barticle}
\endbibitem

\bibitem[\protect\citeauthoryear{Bengochea and Ferraro}{2009}]{bengochea_dark_2009}
\begin{barticle}
\bauthor{\bsnm{Bengochea}, \binits{G.R.}},
\bauthor{\bsnm{Ferraro}, \binits{R.}}:
\batitle{{Dark torsion as the cosmic speed-up}}.
\bjtitle{Phys. Rev. D}
\bvolume{79}(\bissue{12}),
\bfpage{124019}
(\byear{2009})
\doiurl{10.1103/PhysRevD.79.124019} .
Accessed 2023-05-24
\end{barticle}
\endbibitem

\bibitem[\protect\citeauthoryear{Capozziello et~al.}{2017}]{capozziello2017constraining}
\begin{barticle}
\bauthor{\bsnm{Capozziello}, \binits{S.}},
\bauthor{\bsnm{Lambiase}, \binits{G.}},
\bauthor{\bsnm{Saridakis}, \binits{E.}}:
\batitle{{Constraining {\ensuremath{f(T)}} teleparallel gravity by big bang nucleosynthesis: {\ensuremath{f(T)}} cosmology and BBN}}.
\bjtitle{Euro. Phys. J. C}
\bvolume{77},
\bfpage{1}--\blpage{6}
(\byear{2017})
\doiurl{10.1140/epjc/s10052-017-5143-8}
\end{barticle}
\endbibitem

\bibitem[\protect\citeauthoryear{Awad et~al.}{2018}]{Awad_2018}
\begin{barticle}
\bauthor{\bsnm{Awad}, \binits{A.}},
\bauthor{\bsnm{Hanafy}, \binits{W.E.}},
\bauthor{\bsnm{Nashed}, \binits{G.G.L.}},
\bauthor{\bsnm{Odintsov}, \binits{S.D.}},
\bauthor{\bsnm{Oikonomou}, \binits{V.K.}}:
\batitle{{Constant-roll inflation in {\ensuremath{f(T)}} teleparallel gravity}}.
\bjtitle{JCAP}
\bvolume{2018}(\bissue{07}),
\bfpage{026}--\blpage{026}
(\byear{2018})
\doiurl{10.1088/1475-7516/2018/07/026}
\end{barticle}
\endbibitem

\bibitem[\protect\citeauthoryear{Jim{\'{e}}nez et~al.}{2018}]{jimenez2018teleparallel}
\begin{barticle}
\bauthor{\bsnm{Jim{\'{e}}nez}, \binits{J.B.}},
\bauthor{\bsnm{Heisenberg}, \binits{L.}},
\bauthor{\bsnm{Koivisto}, \binits{T.S.}}:
\batitle{{Teleparallel palatini theories}}.
\bjtitle{JCAP}
\bvolume{2018}(\bissue{08}),
\bfpage{039}
(\byear{2018})
\doiurl{10.1088/1475-7516/2018/08/039}
\end{barticle}
\endbibitem

\bibitem[\protect\citeauthoryear{Chaudhary et~al.}{2024}]{chaudhary2024_MCG}
\begin{botherref}
\oauthor{\bsnm{Chaudhary}, \binits{H.}},
\oauthor{\bsnm{Debnath}, \binits{U.}},
\oauthor{\bsnm{Roy}, \binits{T.}},
\oauthor{\bsnm{Maity}, \binits{S.}},
\oauthor{\bsnm{Mustafa}, \binits{G.}},
\oauthor{\bsnm{Arora}, \binits{M.}}:
{Constraints on the parameters of modified Chaplygin-Jacobi and modified Chaplygin-Abel gases in {\ensuremath{f(T)}} gravity}
(2024).
\url{https://arxiv.org/abs/2307.14691}
\end{botherref}
\endbibitem

\bibitem[\protect\citeauthoryear{Cai et~al.}{2016}]{Cai_2016}
\begin{barticle}
\bauthor{\bsnm{Cai}, \binits{Y.-F.}},
\bauthor{\bsnm{Capozziello}, \binits{S.}},
\bauthor{\bsnm{Laurentis}, \binits{M.D.}},
\bauthor{\bsnm{Saridakis}, \binits{E.N.}}:
\batitle{{{\ensuremath{f(T)}} teleparallel gravity and cosmology}}.
\bjtitle{Rept. Prog. Phys.}
\bvolume{79}(\bissue{10}),
\bfpage{106901}
(\byear{2016})
\doiurl{10.1088/0034-4885/79/10/106901}
\end{barticle}
\endbibitem

\bibitem[\protect\citeauthoryear{Nester and Yo}{1999}]{nester1999symmetric}
\begin{botherref}
\oauthor{\bsnm{Nester}, \binits{J.M.}},
\oauthor{\bsnm{Yo}, \binits{H.-J.}}:
{Symmetric teleparallel general relativity}
(1999).
\url{https://arxiv.org/abs/gr-qc/9809049}
\end{botherref}
\endbibitem

\bibitem[\protect\citeauthoryear{Adak and Sert}{2004}]{adak2004solution}
\begin{botherref}
\oauthor{\bsnm{Adak}, \binits{M.}},
\oauthor{\bsnm{Sert}, \binits{O.}}:
{A Solution to Symmetric Teleparallel Gravity}
(2004).
\url{https://arxiv.org/abs/gr-qc/0412007}
\end{botherref}
\endbibitem

\bibitem[\protect\citeauthoryear{Adak et~al.}{2006}]{ADAK_2006}
\begin{barticle}
\bauthor{\bsnm{Adak}, \binits{M.}},
\bauthor{\bsnm{Kalay}, \binits{M.}},
\bauthor{\bsnm{Sert}, \binits{O.}}:
\batitle{{LAGRANGE} {FORMULATION} {OF} {THE} {SYMMETRIC} {TELEPARALLEL} {GRAVITY}}.
\bjtitle{Int. J. Mod. Phys. D}
\bvolume{15}(\bissue{05}),
\bfpage{619}--\blpage{634}
(\byear{2006})
\doiurl{10.1142/s0218271806008474}
\end{barticle}
\endbibitem

\bibitem[\protect\citeauthoryear{Adak et~al.}{2013}]{ADAK_2013}
\begin{barticle}
\bauthor{\bsnm{Adak}, \binits{M.}},
\bauthor{\bsnm{Sert}, \binits{O.}},
\bauthor{\bsnm{Kalay}, \binits{M.}},
\bauthor{\bsnm{Sari}, \binits{M.}}:
\batitle{{SYMMETRIC} {TELEPARALLEL} {GRAVITY}: {SOME} {EXACT} {SOLUTIONS} {AND} {SPINOR} {COUPLINGS}}.
\bjtitle{Int. J. Mod. Phys. A}
\bvolume{28}(\bissue{32}),
\bfpage{1350167}
(\byear{2013})
\doiurl{10.1142/s0217751x13501674}
\end{barticle}
\endbibitem

\bibitem[\protect\citeauthoryear{Jim{\'{e}}nez et~al.}{2018}]{Jim_nez_2018}
\begin{botherref}
\oauthor{\bsnm{Jim{\'{e}}nez}, \binits{J.B.}},
\oauthor{\bsnm{Heisenberg}, \binits{L.}},
\oauthor{\bsnm{Koivisto}, \binits{T.}}:
{Coincident general relativity}.
Phys. Rev. D
\textbf{98}(4)
(2018)
\doiurl{10.1103/physrevd.98.044048}
\end{botherref}
\endbibitem

\bibitem[\protect\citeauthoryear{Dialektopoulos et~al.}{2019}]{Dialektopoulos_2019}
\begin{botherref}
\oauthor{\bsnm{Dialektopoulos}, \binits{K.F.}},
\oauthor{\bsnm{Koivisto}, \binits{T.S.}},
\oauthor{\bsnm{Capozziello}, \binits{S.}}:
{Noether symmetries in symmetric teleparallel cosmology}.
Eur. Phys. J. C
\textbf{79}(7)
(2019)
\doiurl{10.1140/epjc/s10052-019-7106-8}
\end{botherref}
\endbibitem

\bibitem[\protect\citeauthoryear{Jim{\'{e}}nez et~al.}{2020}]{PhysRevD.101.103507}
\begin{barticle}
\bauthor{\bsnm{Jim{\'{e}}nez}, \binits{J.B.}},
\bauthor{\bsnm{Heisenberg}, \binits{L.}},
\bauthor{\bsnm{Koivisto}, \binits{T.}},
\bauthor{\bsnm{Pekar}, \binits{S.}}:
\batitle{{Cosmology in $f(Q)$ geometry}}.
\bjtitle{Phys. Rev. D}
\bvolume{101},
\bfpage{103507}
(\byear{2020})
\doiurl{10.1103/PhysRevD.101.103507}
\end{barticle}
\endbibitem

\bibitem[\protect\citeauthoryear{Bajardi et~al.}{2020}]{bajardi2020bouncing}
\begin{barticle}
\bauthor{\bsnm{Bajardi}, \binits{F.}},
\bauthor{\bsnm{Vernieri}, \binits{D.}},
\bauthor{\bsnm{Capozziello}, \binits{S.}}:
\batitle{{Bouncing cosmology in $f(Q)$ symmetric teleparallel gravity}}.
\bjtitle{Eur. Phys. J. Plus}
\bvolume{135}(\bissue{11}),
\bfpage{1}--\blpage{14}
(\byear{2020})
\doiurl{10.1140/epjp/s13360-020-00918-3}
\end{barticle}
\endbibitem

\bibitem[\protect\citeauthoryear{Mandal et~al.}{2020a}]{PhysRevD.102.124029}
\begin{barticle}
\bauthor{\bsnm{Mandal}, \binits{S.}},
\bauthor{\bsnm{Wang}, \binits{D.}},
\bauthor{\bsnm{Sahoo}, \binits{P.K.}}:
\batitle{{Cosmography in $f(Q)$ gravity}}.
\bjtitle{Phys. Rev. D}
\bvolume{102},
\bfpage{124029}
(\byear{2020})
\doiurl{10.1103/PhysRevD.102.124029}
\end{barticle}
\endbibitem

\bibitem[\protect\citeauthoryear{Mandal et~al.}{2020b}]{mandal2020energy}
\begin{barticle}
\bauthor{\bsnm{Mandal}, \binits{S.}},
\bauthor{\bsnm{Sahoo}, \binits{P.}},
\bauthor{\bsnm{Santos}, \binits{J.}}:
\batitle{{Energy conditions in $f(Q)$ gravity}}.
\bjtitle{Phys. Rev. D}
\bvolume{102}(\bissue{2}),
\bfpage{024057}
(\byear{2020})
\doiurl{10.1103/PhysRevD.102.024057}
\end{barticle}
\endbibitem

\bibitem[\protect\citeauthoryear{Arora and Sahoo}{2022}]{Arora_2022}
\begin{botherref}
\oauthor{\bsnm{Arora}, \binits{S.}},
\oauthor{\bsnm{Sahoo}, \binits{P.K.}}:
{Crossing Phantom Divide in $f(Q)$ Gravity}.
Ann. der Phys.
\textbf{534}(8)
(2022)
\doiurl{10.1002/andp.202200233}
\end{botherref}
\endbibitem

\bibitem[\protect\citeauthoryear{Lu et~al.}{2019}]{lu2019_1906.08920}
\begin{botherref}
\oauthor{\bsnm{Lu}, \binits{J.}},
\oauthor{\bsnm{Zhao}, \binits{X.}},
\oauthor{\bsnm{Chee}, \binits{G.}}:
{Cosmology in symmetric teleparallel gravity and its dynamical system}
(2019).
\url{https://arxiv.org/abs/1906.08920}
\end{botherref}
\endbibitem

\bibitem[\protect\citeauthoryear{Lazkoz et~al.}{2019}]{lazkoz2019observational}
\begin{barticle}
\bauthor{\bsnm{Lazkoz}, \binits{R.}},
\bauthor{\bsnm{Lobo}, \binits{F.S.}},
\bauthor{\bsnm{Ortiz-Ba{\~n}os}, \binits{M.}},
\bauthor{\bsnm{Salzano}, \binits{V.}}:
\batitle{{Observational constraints of $f(Q)$ gravity}}.
\bjtitle{Phys. Rev. D}
\bvolume{100}(\bissue{10}),
\bfpage{104027}
(\byear{2019})
\doiurl{10.1103/PhysRevD.100.104027}
\end{barticle}
\endbibitem

\bibitem[\protect\citeauthoryear{Anagnostopoulos et~al.}{2021}]{anagnostopoulos2021first}
\begin{barticle}
\bauthor{\bsnm{Anagnostopoulos}, \binits{F.K.}},
\bauthor{\bsnm{Basilakos}, \binits{S.}},
\bauthor{\bsnm{Saridakis}, \binits{E.N.}}:
\batitle{{First evidence that non-metricity $f(Q)$ gravity could challenge $\Lambda$CDM}}.
\bjtitle{Phys. Lett. B}
\bvolume{822},
\bfpage{136634}
(\byear{2021})
\doiurl{10.1016/j.physletb.2021.136634}
\end{barticle}
\endbibitem

\bibitem[\protect\citeauthoryear{Narawade and Mishra}{2023}]{Narawade_2023}
\begin{botherref}
\oauthor{\bsnm{Narawade}, \binits{S.A.}},
\oauthor{\bsnm{Mishra}, \binits{B.}}:
{Phantom Cosmological Model with Observational Constraints in $f(Q)$ Gravity}.
Ann. der Phys.
\textbf{535}(5)
(2023)
\doiurl{10.1002/andp.202200626}
\end{botherref}
\endbibitem

\bibitem[\protect\citeauthoryear{Atayde and Frusciante}{2021}]{Atayde_2021}
\begin{botherref}
\oauthor{\bsnm{Atayde}, \binits{L.}},
\oauthor{\bsnm{Frusciante}, \binits{N.}}:
{Can $f(Q)$ gravity challenge $\Lambda$CDM?}
Phys. Rev. D
\textbf{104}(6)
(2021)
\doiurl{10.1103/physrevd.104.064052}
\end{botherref}
\endbibitem

\bibitem[\protect\citeauthoryear{Anagnostopoulos et~al.}{2023}]{anagnostopoulos2023new}
\begin{barticle}
\bauthor{\bsnm{Anagnostopoulos}, \binits{F.K.}},
\bauthor{\bsnm{Gakis}, \binits{V.}},
\bauthor{\bsnm{Saridakis}, \binits{E.N.}},
\bauthor{\bsnm{Basilakos}, \binits{S.}}:
\batitle{{New models and big bang nucleosynthesis constraints in $f(Q)$ gravity}}.
\bjtitle{Eur. Phys. J. C}
\bvolume{83}(\bissue{1}),
\bfpage{58}
(\byear{2023})
\doiurl{10.1140/epjc/s10052-023-11190-x}
\end{barticle}
\endbibitem

\bibitem[\protect\citeauthoryear{Dimakis et~al.}{2021}]{Dimakis_2021}
\begin{barticle}
\bauthor{\bsnm{Dimakis}, \binits{N.}},
\bauthor{\bsnm{Paliathanasis}, \binits{A.}},
\bauthor{\bsnm{Christodoulakis}, \binits{T.}}:
\batitle{{Quantum cosmology in $f(Q)$ theory}}.
\bjtitle{Class. Quantum Grav.}
\bvolume{38}(\bissue{22}),
\bfpage{225003}
(\byear{2021})
\doiurl{10.1088/1361-6382/ac2b09}
\end{barticle}
\endbibitem

\bibitem[\protect\citeauthoryear{Koussour et~al.}{2022}]{koussour2022late}
\begin{barticle}
\bauthor{\bsnm{Koussour}, \binits{M.}},
\bauthor{\bsnm{El~Bourakadi}, \binits{K.}},
\bauthor{\bsnm{Shekh}, \binits{S.}},
\bauthor{\bsnm{Pacif}, \binits{S.}},
\bauthor{\bsnm{Bennai}, \binits{M.}}:
\batitle{{Late-time acceleration in $f(Q)$ gravity: Analysis and constraints in an anisotropic background}}.
\bjtitle{Ann. Phys.}
\bvolume{445},
\bfpage{169092}
(\byear{2022})
\doiurl{10.1016/j.aop.2022.169092}
\end{barticle}
\endbibitem

\bibitem[\protect\citeauthoryear{Solanki et~al.}{2022}]{Solanki_2022}
\begin{barticle}
\bauthor{\bsnm{Solanki}, \binits{R.}},
\bauthor{\bsnm{De}, \binits{A.}},
\bauthor{\bsnm{Sahoo}, \binits{P.K.}}:
\batitle{{Complete dark energy scenario in $f(Q)$ gravity}}.
\bjtitle{Phys. Dark Univ.}
\bvolume{36},
\bfpage{100996}
(\byear{2022})
\doiurl{10.1016/j.dark.2022.100996}
\end{barticle}
\endbibitem

\bibitem[\protect\citeauthoryear{Rana and Sahoo}{2024}]{rana_2024a}
\begin{botherref}
\oauthor{\bsnm{Rana}, \binits{D.S.}},
\oauthor{\bsnm{Sahoo}, \binits{P.K.}}:
Cosmological constraints in symmetric teleparallel gravity with bulk viscosity.
Gen. Relativ. Grav.
\textbf{56}(1572-9532)
(2024)
\doiurl{10.1007/s10714-024-03271-3}
\end{botherref}
\endbibitem

\bibitem[\protect\citeauthoryear{Khyllep et~al.}{2021}]{khyllep2021cosmological}
\begin{barticle}
\bauthor{\bsnm{Khyllep}, \binits{W.}},
\bauthor{\bsnm{Paliathanasis}, \binits{A.}},
\bauthor{\bsnm{Dutta}, \binits{J.}}:
\batitle{{Cosmological solutions and growth index of matter perturbations in $f(Q)$ gravity}}.
\bjtitle{Phys. Rev. D}
\bvolume{103}(\bissue{10}),
\bfpage{103521}
(\byear{2021})
\doiurl{10.1103/physrevd.103.103521}
\end{barticle}
\endbibitem

\bibitem[\protect\citeauthoryear{Khyllep et~al.}{2023}]{PhysRevD.107.044022}
\begin{barticle}
\bauthor{\bsnm{Khyllep}, \binits{W.}},
\bauthor{\bsnm{Dutta}, \binits{J.}},
\bauthor{\bsnm{Saridakis}, \binits{E.N.}},
\bauthor{\bsnm{Yesmakhanova}, \binits{K.}}:
\batitle{{Cosmology in $f(Q)$ gravity: A unified dynamical systems analysis of the background and perturbations}}.
\bjtitle{Phys. Rev. D}
\bvolume{107},
\bfpage{044022}
(\byear{2023})
\doiurl{10.1103/PhysRevD.107.044022}
\end{barticle}
\endbibitem

\bibitem[\protect\citeauthoryear{Narawade et~al.}{2023}]{Narawade_2023_a3}
\begin{barticle}
\bauthor{\bsnm{Narawade}, \binits{S.A.}},
\bauthor{\bsnm{Singh}, \binits{S.P.}},
\bauthor{\bsnm{Mishra}, \binits{B.}}:
\batitle{{Accelerating cosmological models in $f(Q)$ gravity and the phase space analysis}}.
\bjtitle{Phys. Dark Univ.}
\bvolume{42},
\bfpage{101282}
(\byear{2023})
\doiurl{10.1016/j.dark.2023.101282}
\end{barticle}
\endbibitem

\bibitem[\protect\citeauthoryear{Harko et~al.}{2018}]{Harko_2018}
\begin{barticle}
\bauthor{\bsnm{Harko}, \binits{T.}},
\bauthor{\bsnm{Koivisto}, \binits{T.S.}},
\bauthor{\bsnm{Lobo}, \binits{F.S.N.}},
\bauthor{\bsnm{Olmo}, \binits{G.J.}},
\bauthor{\bsnm{Rubiera-Garcia}, \binits{D.}}:
\batitle{{Coupling matter in modified $Q$ gravity}}.
\bjtitle{Phys. Rev. D}
\bvolume{98},
\bfpage{084043}
(\byear{2018})
\doiurl{10.1103/PhysRevD.98.084043}
\end{barticle}
\endbibitem

\bibitem[\protect\citeauthoryear{Xu et~al.}{2019}]{xu2019f}
\begin{barticle}
\bauthor{\bsnm{Xu}, \binits{Y.}},
\bauthor{\bsnm{Li}, \binits{G.}},
\bauthor{\bsnm{Harko}, \binits{T.}},
\bauthor{\bsnm{Liang}, \binits{S.-D.}}:
\batitle{{$f(Q, T)$ gravity}}.
\bjtitle{Eur. Phys. J. C}
\bvolume{79}(\bissue{8}),
\bfpage{1}--\blpage{19}
(\byear{2019})
\doiurl{10.1140/epjc/s10052-019-7207-4}
\end{barticle}
\endbibitem

\bibitem[\protect\citeauthoryear{Pradhan et~al.}{2023}]{Pradhan_2023}
\begin{barticle}
\bauthor{\bsnm{Pradhan}, \binits{A.}},
\bauthor{\bsnm{Goswami}, \binits{G.}},
\bauthor{\bsnm{Beesham}, \binits{A.}}:
\batitle{{The reconstruction of constant jerk parameter with $f(R,T)$ gravity}}.
\bjtitle{JHEAp}
\bvolume{38},
\bfpage{12}--\blpage{21}
(\byear{2023})
\doiurl{10.1016/j.jheap.2023.03.001}
\end{barticle}
\endbibitem

\bibitem[\protect\citeauthoryear{Sahni et~al.}{2003}]{Sahni_2003}
\begin{barticle}
\bauthor{\bsnm{Sahni}, \binits{V.}},
\bauthor{\bsnm{Saini}, \binits{T.D.}},
\bauthor{\bsnm{Starobinsky}, \binits{A.A.}},
\bauthor{\bsnm{Alam}, \binits{U.}}:
\batitle{Statefinder-a new geometrical diagnostic of dark energy}.
\bjtitle{JETP Lett.}
\bvolume{77}(\bissue{5}),
\bfpage{201}--\blpage{206}
(\byear{2003})
\doiurl{10.1134/1.1574831}
\end{barticle}
\endbibitem

\bibitem[\protect\citeauthoryear{Sharov and Vasiliev}{2018}]{Sharov_2018}
\begin{botherref}
\oauthor{\bsnm{Sharov}, \binits{G.S.}},
\oauthor{\bsnm{Vasiliev}, \binits{V.O.}}:
{How predictions of cosmological models depend on Hubble parameter data sets}.
Mathematical Modelling and Geometry
\textbf{6}(1)
(2018)
\doiurl{10.26456/mmg/2018-611}
\end{botherref}
\endbibitem

\bibitem[\protect\citeauthoryear{Scolnic et~al.}{2018}]{Scolnic_2018}
\begin{barticle}
\bauthor{\bsnm{Scolnic}, \binits{D.M.}}, \betal:
\batitle{{The Complete Light-curve Sample of Spectroscopically Confirmed SNe Ia from Pan-STARRS1 and Cosmological Constraints from the Combined Pantheon Sample}}.
\bjtitle{Astrophys. J.}
\bvolume{859}(\bissue{2}),
\bfpage{101}
(\byear{2018})
\doiurl{10.3847/1538-4357/aab9bb}
\end{barticle}
\endbibitem

\bibitem[\protect\citeauthoryear{{Di Valentino} et~al.}{2016}]{DIVALENTINO2016242}
\begin{barticle}
\bauthor{\bsnm{{Di Valentino}}, \binits{E.}},
\bauthor{\bsnm{Melchiorri}, \binits{A.}},
\bauthor{\bsnm{Silk}, \binits{J.}}:
\batitle{Reconciling planck with the local value of $h_0$ in extended parameter space}.
\bjtitle{Phys. Lett. B}
\bvolume{761},
\bfpage{242}--\blpage{246}
(\byear{2016})
\doiurl{10.1016/j.physletb.2016.08.043}
\end{barticle}
\endbibitem

\bibitem[\protect\citeauthoryear{Amanullah et~al.}{2010}]{amanullah2010spectra}
\begin{barticle}
\bauthor{\bsnm{Amanullah}, \binits{R.}}, \betal:
\batitle{{Spectra and Hubble Space Telescope light curves of six type Ia supernovae at {\ensuremath{0.511 < z < 1.12}} and the Union2 compilation}}.
\bjtitle{Astrophys. J.}
\bvolume{716}(\bissue{1}),
\bfpage{712}
(\byear{2010})
\doiurl{10.1088/0004-637X/716/1/712}
\end{barticle}
\endbibitem

\bibitem[\protect\citeauthoryear{Planck~Collaboration et~al.}{2020}]{planck_2018_results}
\begin{barticle}
\bauthor{\bsnm{Planck~Collaboration}, \binits{N.} \bsuffix{Aghanim}},
\bauthor{\bsnm{Akrami}, \binits{Y.}}, \betal:
\batitle{{Planck2018 results: VI. Cosmological parameters}}.
\bjtitle{Astron. Astrophys.}
\bvolume{641},
\bfpage{6}
(\byear{2020})
\doiurl{10.1051/0004-6361/201833910}
\end{barticle}
\endbibitem

\bibitem[\protect\citeauthoryear{Ayuso et~al.}{2021}]{PhysRevD.103.063505}
\begin{barticle}
\bauthor{\bsnm{Ayuso}, \binits{I.}},
\bauthor{\bsnm{Lazkoz}, \binits{R.}},
\bauthor{\bsnm{Salzano}, \binits{V.}}:
\batitle{{Observational constraints on cosmological solutions of $f(Q)$ theories}}.
\bjtitle{Phys. Rev. D}
\bvolume{103},
\bfpage{063505}
(\byear{2021})
\doiurl{10.1103/PhysRevD.103.063505}
\end{barticle}
\endbibitem

\bibitem[\protect\citeauthoryear{Zhang et~al.}{2014}]{Cong_2014h}
\begin{botherref}
\oauthor{\bsnm{Zhang}, \binits{C.}},
\oauthor{\bsnm{Zhang}, \binits{H.}},
\oauthor{\bsnm{Yuan}, \binits{S.}},
\oauthor{\bsnm{Liu}, \binits{S.}},
\oauthor{\bsnm{Zhang}, \binits{T.-J.}},
\oauthor{\bsnm{Sun}, \binits{Y.-C.}}:
{Four New Observational $H(z)$ Data From Luminous Red Galaxies of Sloan Digital Sky Survey Data Release Seven}.
Astron. Astrophys.
\textbf{14}
(2014)
\doiurl{10.1088/1674-4527/14/10/002}
\end{botherref}
\endbibitem

\bibitem[\protect\citeauthoryear{Moresco et~al.}{2016}]{Moresco_2016}
\begin{barticle}
\bauthor{\bsnm{Moresco}, \binits{M.}}, \betal:
\batitle{{A 6\% measurement of the Hubble parameter at $z \sim 0.45$: direct evidence of the epoch of cosmic re-acceleration}}.
\bjtitle{JCAP}
\bvolume{2016}(\bissue{05}),
\bfpage{014}--\blpage{014}
(\byear{2016})
\doiurl{10.1088/1475-7516/2016/05/014}
\end{barticle}
\endbibitem

\bibitem[\protect\citeauthoryear{Stern et~al.}{2010}]{Daniel_Stern_2010}
\begin{barticle}
\bauthor{\bsnm{Stern}, \binits{D.}},
\bauthor{\bsnm{Jimenez}, \binits{R.}},
\bauthor{\bsnm{Verde}, \binits{L.}},
\bauthor{\bsnm{Kamionkowski}, \binits{M.}},
\bauthor{\bsnm{Stanford}, \binits{S.A.}}:
\batitle{{Cosmic chronometers: constraining the equation of state of dark energy. I: $H(z)$ measurements}}.
\bjtitle{JCAP}
\bvolume{2010}(\bissue{02}),
\bfpage{008}--\blpage{008}
(\byear{2010})
\doiurl{10.1088/1475-7516/2010/02/008}
\end{barticle}
\endbibitem

\bibitem[\protect\citeauthoryear{Moresco et~al.}{2012}]{M_Moresco_2012}
\begin{barticle}
\bauthor{\bsnm{Moresco}, \binits{M.}},
\bauthor{\bsnm{Cimatti}, \binits{A.}},
\bauthor{\bsnm{Jimenez}, \binits{R.}}, \betal:
\batitle{{Improved constraints on the expansion rate of the Universe up to $z \sim 1.1$ from the spectroscopic evolution of cosmic chronometers}}.
\bjtitle{JCAP}
\bvolume{2012}(\bissue{08}),
\bfpage{006}--\blpage{006}
(\byear{2012})
\doiurl{10.1088/1475-7516/2012/08/006}
\end{barticle}
\endbibitem

\bibitem[\protect\citeauthoryear{Moresco}{2015}]{Moresco_2015}
\begin{barticle}
\bauthor{\bsnm{Moresco}, \binits{M.}}:
\batitle{{Raising the bar: new constraints on the Hubble parameter with cosmic chronometers at $z \sim 2$}}.
\bjtitle{MNRAS: Lett.}
\bvolume{450}(\bissue{1}),
\bfpage{16}--\blpage{20}
(\byear{2015})
\doiurl{10.1093/mnrasl/slv037}
\end{barticle}
\endbibitem

\bibitem[\protect\citeauthoryear{Ratsimbazafy et~al.}{2017}]{Ratsimbazafy_2017}
\begin{barticle}
\bauthor{\bsnm{Ratsimbazafy}, \binits{A.L.}}, \betal:
\batitle{{Age-dating luminous red galaxies observed with the Southern African Large Telescope}}.
\bjtitle{MNRAS}
\bvolume{467},
\bfpage{3239}--\blpage{3254}
(\byear{2017})
\doiurl{10.1093/mnras/stx301}
\end{barticle}
\endbibitem

\bibitem[\protect\citeauthoryear{Hicken et~al.}{2009}]{Hicken_2009}
\begin{barticle}
\bauthor{\bsnm{Hicken}, \binits{M.}}, \betal:
\batitle{{CfA3: 185 TYPE Ia SUPERNOVA LIGHT CURVES FROM THE CfA}}.
\bjtitle{Astrophys. J.}
\bvolume{700}(\bissue{1}),
\bfpage{331}
(\byear{2009})
\doiurl{10.1088/0004-637X/700/1/331}
\end{barticle}
\endbibitem

\bibitem[\protect\citeauthoryear{Sako et~al.}{2018}]{Sako_2018}
\begin{barticle}
\bauthor{\bsnm{Sako}, \binits{M.}}, \betal:
\batitle{{The Data Release of the Sloan Digital Sky Survey-II Supernova Survey}}.
\bjtitle{Publications of the Astronomical Society of the Pacific}
\bvolume{130}(\bissue{988}),
\bfpage{064002}
(\byear{2018})
\doiurl{10.1088/1538-3873/aab4e0}
\end{barticle}
\endbibitem

\bibitem[\protect\citeauthoryear{Guy et~al.}{2010}]{Guy_2010}
\begin{barticle}
\bauthor{\bsnm{Guy}, \binits{J.}}, \betal:
\batitle{{The Supernova Legacy Survey 3-year sample: Type Ia supernovae photometric distances and cosmological constraints}}.
\bjtitle{AA}
\bvolume{523},
\bfpage{7}
(\byear{2010})
\doiurl{10.1051/0004-6361/201014468}
\end{barticle}
\endbibitem

\bibitem[\protect\citeauthoryear{Narayan et~al.}{2016}]{Narayan_2016}
\begin{barticle}
\bauthor{\bsnm{Narayan}, \binits{G.}}, \betal:
\batitle{{LIGHT CURVES OF 213 TYPE Ia SUPERNOVAE FROM THE ESSENCE SURVEY}}.
\bjtitle{Astrophys. J. Suppl.}
\bvolume{224}(\bissue{1}),
\bfpage{3}
(\byear{2016})
\doiurl{10.3847/0067-0049/224/1/3}
\end{barticle}
\endbibitem

\bibitem[\protect\citeauthoryear{Contreras et~al.}{2010}]{Contreras_2010}
\begin{barticle}
\bauthor{\bsnm{Contreras}, \binits{C.}}, \betal:
\batitle{{The Carnegie Supernova Project: First Photometry Data Release of Low-Redshift Type Ia Supernovae}}.
\bjtitle{Astron. J.}
\bvolume{139}(\bissue{2}),
\bfpage{519}
(\byear{2010})
\doiurl{10.1088/0004-6256/139/2/519}
\end{barticle}
\endbibitem

\bibitem[\protect\citeauthoryear{Graur et~al.}{2014}]{Graur_2014}
\begin{barticle}
\bauthor{\bsnm{Graur}, \binits{O.}}, \betal:
\batitle{{TYPE-Ia SUPERNOVA RATES TO REDSHIFT 2.4 FROM CLASH: THE CLUSTER LENSING AND SUPERNOVA SURVEY WITH HUBBLE}}.
\bjtitle{Astrophys. J.}
\bvolume{783}(\bissue{1}),
\bfpage{28}
(\byear{2014})
\doiurl{10.1088/0004-637X/783/1/28}
\end{barticle}
\endbibitem

\bibitem[\protect\citeauthoryear{Riess et~al.}{2018}]{Riess_2018}
\begin{barticle}
\bauthor{\bsnm{Riess}, \binits{A.G.}}, \betal:
\batitle{{Type Ia Supernova Distances at Redshift $> 1.5$ from the Hubble Space Telescope Multi-cycle Treasury Programs: The Early Expansion Rate}}.
\bjtitle{Astrophys. J.}
\bvolume{853}(\bissue{2}),
\bfpage{126}
(\byear{2018})
\doiurl{10.3847/1538-4357/aaa5a9}
\end{barticle}
\endbibitem

\bibitem[\protect\citeauthoryear{Riess et~al.}{2007}]{Riess_2007}
\begin{barticle}
\bauthor{\bsnm{Riess}, \binits{A.G.}}, \betal:
\batitle{{New Hubble Space Telescope Discoveries of Type Ia Supernovae at $z \geq 1$: Narrowing Constraints on the Early Behavior of Dark Energy}}.
\bjtitle{Astrophys. J.}
\bvolume{659}(\bissue{1}),
\bfpage{98}
(\byear{2007})
\doiurl{10.1086/510378}
\end{barticle}
\endbibitem

\bibitem[\protect\citeauthoryear{Asvesta et~al.}{2022}]{Asvesta_2022}
\begin{barticle}
\bauthor{\bsnm{Asvesta}, \binits{K.}},
\bauthor{\bsnm{Kazantzidis}, \binits{L.}},
\bauthor{\bsnm{Perivolaropoulos}, \binits{L.}},
\bauthor{\bsnm{Tsagas}, \binits{C.G.}}:
\batitle{{Observational constraints on the deceleration parameter in a tilted universe}}.
\bjtitle{MNRAS}
\bvolume{513}(\bissue{2}),
\bfpage{2394}--\blpage{2406}
(\byear{2022})
\doiurl{10.1093/mnras/stac922}
\end{barticle}
\endbibitem

\bibitem[\protect\citeauthoryear{Riess et~al.}{2021}]{Riess_2021}
\begin{barticle}
\bauthor{\bsnm{Riess}, \binits{A.G.}},
\bauthor{\bsnm{Casertano}, \binits{S.}},
\bauthor{\bsnm{Yuan}, \binits{W.}},
\bauthor{\bsnm{Bowers}, \binits{J.B.}},
\bauthor{\bsnm{Macri}, \binits{L.}},
\bauthor{\bsnm{Zinn}, \binits{J.C.}},
\bauthor{\bsnm{Scolnic}, \binits{D.}}:
\batitle{{Cosmic Distances Calibrated to 1\% Precision with Gaia EDR3 Parallaxes and Hubble Space Telescope Photometry of 75 Milky Way Cepheids Confirm Tension with $\Lambda$CDM}}.
\bjtitle{Astrophys. J. Lett.}
\bvolume{908}(\bissue{1}),
\bfpage{6}
(\byear{2021})
\doiurl{10.3847/2041-8213/abdbaf}
\end{barticle}
\endbibitem

\bibitem[\protect\citeauthoryear{Cao and Ratra}{2023}]{Cao_2023}
\begin{botherref}
\oauthor{\bsnm{Cao}, \binits{S.}},
\oauthor{\bsnm{Ratra}, \binits{B.}}:
{${H}_{0}=69.8\ifmmode\pm\else\textpm\fi{}1.3\text{ }\text{ }\mathrm{km}\text{ }{\mathrm{s}}^{\ensuremath{-}1}\text{ }{\mathrm{Mpc}}^{\ensuremath{-}1}$, ${\mathrm{\ensuremath{\Omega}}}_{m0}=0.288\ifmmode\pm\else\textpm\fi{}0.017$, and other constraints from lower-redshift, non-CMB, expansion-rate data}.
Phys. Rev. D
\textbf{107}(10)
(2023)
\doiurl{10.1103/physrevd.107.103521}
\end{botherref}
\endbibitem

\bibitem[\protect\citeauthoryear{Cao and Ratra}{2022}]{Cao_2022}
\begin{barticle}
\bauthor{\bsnm{Cao}, \binits{S.}},
\bauthor{\bsnm{Ratra}, \binits{B.}}:
\batitle{{Using lower-redshift, non-CMB, data to constrain the Hubble constant and other cosmological parameters}}.
\bjtitle{Mon. Not. Roy. Astron. Soc.}
(\byear{2022})
\doiurl{10.1093/mnras/stac1184}
\end{barticle}
\endbibitem

\bibitem[\protect\citeauthoryear{Domínguez et~al.}{2019}]{Dom_nguez_2019}
\begin{barticle}
\bauthor{\bsnm{Domínguez}, \binits{A.}}, \betal:
\batitle{{A New Measurement of the Hubble Constant and Matter Content of the Universe Using Extragalactic Background Light $\gamma$-Ray Attenuation}}.
\bjtitle{Astrophys. J.}
\bvolume{885}(\bissue{2}),
\bfpage{137}
(\byear{2019})
\doiurl{10.3847/1538-4357/ab4a0e}
\end{barticle}
\endbibitem

\bibitem[\protect\citeauthoryear{Park and Ratra}{2020}]{Park_2020}
\begin{botherref}
\oauthor{\bsnm{Park}, \binits{C.-G.}},
\oauthor{\bsnm{Ratra}, \binits{B.}}:
{Using SPTpol, Planck 2015, and non-CMB data to constrain tilted spatially-flat and untilted non-flat $\Lambda$CDM, XCDM, and $\phi$CDM dark energy inflation cosmologies}.
Phys. Rev. D
\textbf{101}(8)
(2020)
\doiurl{10.1103/physrevd.101.083508}
\end{botherref}
\endbibitem

\bibitem[\protect\citeauthoryear{Lin and Ishak}{2021}]{Lin_2021}
\begin{barticle}
\bauthor{\bsnm{Lin}, \binits{W.}},
\bauthor{\bsnm{Ishak}, \binits{M.}}:
\batitle{{A Bayesian interpretation of inconsistency measures in cosmology}}.
\bjtitle{JCAP}
\bvolume{2021}(\bissue{05}),
\bfpage{009}
(\byear{2021})
\doiurl{10.1088/1475-7516/2021/05/009}
\end{barticle}
\endbibitem

\bibitem[\protect\citeauthoryear{Freedman et~al.}{2020}]{Freedman_2020}
\begin{barticle}
\bauthor{\bsnm{Freedman}, \binits{W.L.}},
\bauthor{\bsnm{Madore}, \binits{B.F.}},
\bauthor{\bsnm{Hoyt}, \binits{T.}},
\bauthor{\bsnm{Jang}, \binits{I.S.}},
\bauthor{\bsnm{Beaton}, \binits{R.}},
\bauthor{\bsnm{Lee}, \binits{M.G.}},
\bauthor{\bsnm{Monson}, \binits{A.}},
\bauthor{\bsnm{Neeley}, \binits{J.}},
\bauthor{\bsnm{Rich}, \binits{J.}}:
\batitle{{Calibration of the Tip of the Red Giant Branch}}.
\bjtitle{Astrophys. J.}
\bvolume{891}(\bissue{1}),
\bfpage{57}
(\byear{2020})
\doiurl{10.3847/1538-4357/ab7339}
\end{barticle}
\endbibitem

\bibitem[\protect\citeauthoryear{Birrer et~al.}{2020}]{Birrer_2020}
\begin{barticle}
\bauthor{\bsnm{Birrer}, \binits{S.}},
\bauthor{\bsnm{Shajib}, \binits{A.J.}},
\bauthor{\bsnm{Galan}, \binits{A.}},
\bauthor{\bsnm{Millon}, \binits{M.}}, \betal:
\batitle{{TDCOSMO: IV. Hierarchical time-delay cosmography – joint inference of the Hubble constant and galaxy density profiles}}.
\bjtitle{Astron. Astrophys.}
\bvolume{643},
\bfpage{165}
(\byear{2020})
\doiurl{10.1051/0004-6361/202038861}
\end{barticle}
\endbibitem

\bibitem[\protect\citeauthoryear{Boruah et~al.}{2021}]{Boruah_2021}
\begin{barticle}
\bauthor{\bsnm{Boruah}, \binits{S.S.}},
\bauthor{\bsnm{Hudson}, \binits{M.J.}},
\bauthor{\bsnm{Lavaux}, \binits{G.}}:
\batitle{{Peculiar velocities in the local Universe: comparison of different models and the implications for $H_{0}$ and dark matter}}.
\bjtitle{MNRAS}
\bvolume{507}(\bissue{2}),
\bfpage{2697}--\blpage{2713}
(\byear{2021})
\doiurl{10.1093/mnras/stab2320}
\end{barticle}
\endbibitem

\bibitem[\protect\citeauthoryear{Freedman}{2021}]{Freedman_2021}
\begin{barticle}
\bauthor{\bsnm{Freedman}, \binits{W.L.}}:
\batitle{{Measurements of the Hubble Constant: Tensions in Perspective*}}.
\bjtitle{Astrophys. J.}
\bvolume{919}(\bissue{1}),
\bfpage{16}
(\byear{2021})
\doiurl{10.3847/1538-4357/ac0e95}
\end{barticle}
\endbibitem

\bibitem[\protect\citeauthoryear{Wu et~al.}{2022}]{Wu_2022}
\begin{barticle}
\bauthor{\bsnm{Wu}, \binits{Q.}},
\bauthor{\bsnm{Zhang}, \binits{G.-Q.}},
\bauthor{\bsnm{Wang}, \binits{F.-Y.}}:
\batitle{{An 8\% determination of the Hubble constant from localized fast radio bursts}}.
\bjtitle{MNRAS: Lett.}
\bvolume{515}(\bissue{1}),
\bfpage{1}--\blpage{5}
(\byear{2022})
\doiurl{10.1093/mnrasl/slac022}
\end{barticle}
\endbibitem

\bibitem[\protect\citeauthoryear{Raychaudhuri}{1955}]{PhysRev.98.1123}
\begin{barticle}
\bauthor{\bsnm{Raychaudhuri}, \binits{A.}}:
\batitle{{Relativistic Cosmology. I}}.
\bjtitle{Phys. Rev.}
\bvolume{98},
\bfpage{1123}--\blpage{1126}
(\byear{1955})
\doiurl{10.1103/PhysRev.98.1123}
\end{barticle}
\endbibitem

\bibitem[\protect\citeauthoryear{Nojiri and Odintsov}{2007}]{NOJIRI_2007}
\begin{barticle}
\bauthor{\bsnm{Nojiri}, \binits{S.}},
\bauthor{\bsnm{Odintsov}, \binits{S.D.}}:
\batitle{{INTRODUCTION TO MODIFIED GRAVITY AND GRAVITATIONAL ALTERNATIVE FOR DARK ENERGY}}.
\bjtitle{Int. J. Geom. Methods Mod. Phys.}
\bvolume{04}(\bissue{01}),
\bfpage{115}--\blpage{145}
(\byear{2007})
\doiurl{10.1142/s0219887807001928}
\end{barticle}
\endbibitem

\end{thebibliography}

\end{document}